\documentclass[a4paper,11pt]{scrartcl}
\usepackage{amsmath}
\usepackage{amsfonts}
\usepackage{amssymb,mathrsfs}
\usepackage{t1enc}
\usepackage[latin1]{inputenc}
\usepackage[english]{babel}
\usepackage[numbers]{natbib}
 \usepackage{mathtools} 
 
\usepackage{amssymb}
\usepackage[final]{pdfpages}
\usepackage{caption}
\usepackage{url}

\usepackage[colorlinks=false, linktocpage=true,backref,pagebackref,hidelinks]{hyperref}

\newcommand{\R}{\mathbb R}

\newcommand{\beq}{\begin{equation}}
\newcommand{\eeq}{\end{equation}}
\newcommand{\beqarr}{\begin{eqnarray}}
\newcommand{\eeqarr}{\end{eqnarray}}
\newcommand{\beqa}{\begin{eqnarray*}}
\newcommand{\eeqa}{\end{eqnarray*}}
\begin{document}
\thispagestyle{empty}

\title{   \normalsize 
Bridging the gap between dark matter and MOND by a relativistc scalar field approach}

\author{\normalsize Erhard Scholz\footnote{University of Wuppertal, School of  Math./Natural Sciences, D-42117 Wuppertal, Germany \quad  \\
escholz@uni-wuppertal.de \hfill{\scriptsize  ORCID 0000-0001-6741-808X} }
}
\date{\small \today } 
\maketitle
\small 
\begin{abstract}
A Lagrangian model for a  general relativistic  scalar field,  formulated in the  framework of integrable Weyl geometry, is studied. Under the present  assumptions it modifies the light cone structure and induces MOND-like dynamics  in the weak field  approximation of the Einstein frame (gauge).  The Lagrangian  contains   a  Bekenstein-type (``aquadratic'')  term and  a  second order term  generating additional mass energy for the scalar field. Both are  switched on only if the  the scalar field gradient is spacelike and below a MOND-typical  threshold, like in the superfluid model of  Berezhiani/Khoury.   In the 
 weak field limit the Bekenstein term implies  a deep MOND equation for the scalar field and leads to   MOND\-ian free fall trajectories. 
The Lagrangian mass term induces non-negligible energy and pressures of the scalar field with the respective consequences for gravitational light deflection. 
\end{abstract}

{\textbf{Keywords}: scalar field, scale invariance, MOND, radial-acceleration-relation, galaxy, cluster
\setcounter{tocdepth}{2}
\tableofcontents

\vspace*{3em}
\setcounter{section}{0}
\section{\small Introduction}
\subsection{\small Basic idea}
This paper shows that  under specific conditions for its Lagrangian  a gravitationally coupled scalar field can lead to a  general relativistic underpinning of 
 modified Newtonian dynamics (MOND) \citep{Famaey/McGaugh:MOND}. The  model  takes up  Milgrom's and 
  Bekenstein's idea of an ``aquadratic'' Lagrangian \citep{Bekenstein/Milgrom:1984,Bekenstein:2006,Bekenstein:2010},  but modifies it  in several respects: \\
  (i) It works  in the framework of (integrable) Weyl geometry with scale covariant fields,   scale invariant expressions for the  Lagrangian densities and a gravitationlly coupled scalar field. 
   \\ (ii) A mass generating term is assumed for the scalar field (in addition to the Bekenstein term). (iii) The Bekenstein and mass terms  of the scalar field Lagrangian   are assumed to be active only for regions in which the  field gradient is spacelike  and below a MOND-typical threshold. The physical reason for this could be  a destabilization of the substrate underlying the field dynamics for field gradients above a critical threshold  like in Berezhiani/Khoury's superfluid approach \citep{Berezhiani/Khoury:2015,Berezhiani/Khoury:2016}. 

Points (i) and (ii)  change the character of the scalar field in comparison with Bekenstein's. It can  no longer be considered exclusively as an enhancement of the gravitational structure  like the scalar fields  of Jordan-Brans-Dicke  (JBD) type. It rather carries also features of some kind of dark matter sui generis, although no particle ontology is assumed. The mass term of the scalar field  is its critical pivot; it must not destroy the scalar field equation dominated by the Bekenstein term and ought to have a ``reasonable'' Hilbert tensor, i.e, one that is essentially divergence free (on shell of the Einstein equation) and is strong enough for modifying the gravitational field  such that test particles and light ray  structure conform essentially to the expectations  of  MOND. 
 For the sake of a coherently scale covariant  approach the  methods and symbolism of Weyl geometric field theory are used. 
 
\subsection{\small Other relativistic generalizations of MOND \label{subsection other relativistic MOND}}
Diverse attempts at general relativistic enhancements of Milgrom's  modified Newtonian dynamics have been made  during the last forty years. The first step in this direction  has  already been made  in the appendix of the early paper by Bekenstein and Milgrom \citep{Bekenstein/Milgrom:1984}. It is  based on a JBD-like scalar field with 
 an ``aquadratic'' kinetic term and two conformally related Riemannian metrics. This implied, however, a much too small
gravitational light deflection. Often  
the  reason is seen in the fact that  conformal transformations do not change  the light cone structure;  the extremely small energy tensor of the Bekenstein scalar field is  mentioned only in  side remarks.  Follow up papers did not try to  enhance  the mass-energy of the scalar field, but chose the path of   drastically  modifying the metric, for example by  a pair of two ``disformal'' metrics $g_{\mu\nu}$  and $\tilde{g}_{\mu\nu}$,  related to each other by a dynamical structure involving a  timelike vector field $A^{\mu}$ in addition to a scalar field $\phi$. 
 This approach became known under the acronym TeVeS  \citep{Sanders:1997,Bekenstein:2004,Bekenstein:2006,Bekenstein:2010}.
 
For nearly two decades TeVeS   was the main candidate for a relativistic generalization of MOND.  
 Its two metric hypothesis implied, however,   a  subluminal propagation speed for tensorial modes of gravitational waves. This  led to its empirical devalidation (``refutation'')  by the observation  of a wave event (GW170817) in August 2017; a comparison with gamma ray signals clearly indicated  that gravitational waves propagate on the light cone structure of spacetime. 
  
 In the following years two  research programs, both in different ways modifying TeVeS,  gained attention, the BIMOND (``bimetric MOND'') program of Milgrom, initiated in 2009 (i.e., before GW170817) \citep{Milgrom:2009,Milgrom:2010bimetric,Milgrom:2022} and the RMOND approach (`` new relativistic MOND'')  by Skordis and  Z{\l}o\'snik published in 2019 \citep{Skordis/Zlosnik:2019,Skordis/Zlosnik:2021}.  They still  play the central role in the present discourse in the MOND community on relativistic generalizations of their framework.  
Other attempts at deriving MOND in a general relativistic framework have been made by  Berezhiani/Khoury 
using a superfluid hypothesis \citep{Berezhiani/Khoury:2015,Berezhiani/Khoury:2016}, or Hossenfelder et al. starting from Linde's  ``emergent gravity'' approach \citep{Hossenfelder:2017,Hossenfelder/Mistele:2018}. Attempts at  generalizing  MOND in a  Weyl geometric framework, different from the one in the present paper, have been proposed by Maeder   \citep{Maeder:2017,Maeder:2023}  and  Harko et al.  \citep{Burikham/Harko-ea:2023}.  

\subsection{\small Organisation of the paper}
In section \ref{section fundamentals} the fundamentals of the present model are being laid,  with its peculiar Lagrangian for a gravitationally coupled scalar field $\phi$ and with the resulting dynamical equations. 
The metric satisfies an Einstein equation  with an energy tensor of the scalar field in   addition to the baryonic  source term. The scalar field satisfies a relativistic generalization of the non-linear Poisson equation known from MOND;  here it is called the {\em Milgrom equation} (section \ref{subsection dynamical equations}). 
The Lagrangian foresees a screening of the scalar field dynamics above a threshold for the field's gradient, which leaves it   effective in  weak field constellations  only.   The dominant terms under such conditions  are studied in subsection \ref{subsection weak fields}. In the flat space limit the  Newton approximation of relativistic gravity acquires a   potential term $\Phi^{(\phi)}$  due to the scalar field  in addition to the baryonic potential $\Phi^{(bar)}$. The  scalar field 
satisfies a non-linear Poisson (deep MOND)  equation  (section \ref{subsection flat space limit}). Both together are called {\em Newton-Milgrom approximation}.   The flat space potential $\Phi^{(\phi)}$ allows to draw  approximate inferences on the relativistic scalar field $\phi$ (section \ref{reverse Milgrom approximation}). On the other hand, calculations in the MOND-algorithm with a  specific type of MONDian ``interpolation'' functions become possible  in the flat space approximation. They are derived in subsection \ref{subsection MOND interpolation functions}.

The next two sections treat special cases. Section \ref{section central symmetry} focusses on centrally symmetric constellations in baryonic vacuum, which are  important for the gravitational environment of  stars and of simplified (roundish)  model investigations of galaxies. In the inner region the dynamics of the scalar field  is  suppressed, the Lagrangian reduces to Einstein gravity and leads back to the Schwarzschild solution (section \ref{subsection Schwarzschild}). 
In the outer region, with an effective  scalar field,  the solution of the Milgrom equation contributes  an energy tensor to the right hand side of the Einstein equation and shifts the metric away from   the Schwarzschild case. This effect is  discussed quantitatively in subsection \ref{subsection MOND modification of Schwarzschild}. 

Section \ref{section galaxies} deals with galactic dynamics.  At first it addresses the  radial accelerations of galaxies derived in our model. They  are compared with the radial acceleration function empirically determined by McGaugh et al.  \citep{{McGaugh-et-al:2016}}   in the outskirts of rotationally supported galaxies (section \ref{subsection radial acc relation}).
A look at recent data of radial velocities in the Milky Way  follows; they have given rise to serious doubts as to the feasibility of MOND in general \citep{Ou-et-al:2023,Chan-et-al:2023}. It may therefore come as a surprise that the total accelerations of our model derived from the baryonic density used in  \citep{Ou-et-al:2023}  fit the data quite well,  certainly better than MOND with the standard interpolation functions and even better than with the empirical function of McGaugh et al.  (section \ref{subsection Milky Way}). Finally  a glance is shed at the Newtonian mass equivalent of the scalar field energy in the environment of approximately round galaxies (section \ref{subsection halo of galaxies}, \ref{subsection criticism}). 

Open problems are discussed in section \ref{section problems}. The scalar field energy tensor of the scalar field has strong relativistic pressure terms with considerable impact on the light deflection. 
This may be a critical feature for empirical tests of the approach  (section \ref{subsection light deflection}). Moreover, the scalar field halos of galaxies  add mass  to the total dynamical mass of clusters. Under  plausible assumptions   this may become a clue for the dark matter problem in galaxy clusters (section \ref{subsection clusters}). On the level of large scale cosmology, on the other hand, the present approach does not change the outlook of Einstein gravity; this may be seen as a plus or as a weak spot (section \ref{subsection cosmology}). A short look at criticism of MOND in general and the claim of an inconsistency of MOND with recent data on radial velocities in the Milky Way is discussed and rebutted in subsection \ref{subsection criticism}. 

The paper is rounded off by a resum\'ee (section \ref{section resumee}) and an  appendix containing some technical explanations (\ref{section appendix})

\vspace{2em}
\section{\small Fundamentals \label{section fundamentals}}
\subsection{\small Lagrangian \label{subsection Lagrangian}}
We start from a scale and diffeomorphism invariant Lagrangian in dimension $n=4$ 
\beq
\mathcal{L} = \mathcal{L}_H(g,\varphi,\phi) + \mathcal{L}_{\phi}(g,\varphi,\phi) + [\mathcal{L}_{bar}(g_E,Y)]  \, \label{eq Lagrangian general form} 
\eeq
 with dynamical fields $g, \phi$, $Y$ and densities
\[
\mathcal{L}_X = L_X \sqrt{|g|} \qquad  \mbox{(scale weights $w(L_X)=-4$)} 
\]
for all contributions of type $X$.
 Here $g$ stands for the Riemannian component of a Weylian metric $[(g,\varphi)]$ with scale connection  $\varphi$ which is assumed to be  integrable, i.e., non-dynamical (see appendix \ref{appendix Weyl geometry}).  $\phi$ is a scale covariant scalar field of weight $w(\phi)= -1$, 
  $\mathcal{L}_H$ is the gravitational Hilbert term with non-minimally coupled  scalar field 
\beq   
   \mathcal{L}_H = \frac{(\hbar c)^{-1}}{2} (\xi \phi)^2 R(g,\varphi) \, \sqrt{|g|} \, . \label{eq Hilbert Lagrangian} \qquad  
\eeq
 $R(g,\varphi)$ is the Weyl-geometric scalar curvature (of weight $-2$).\footnote{The factors $(\hbar c)^{-k}$ ($k=1,3$) here and below  ensure that in Einstein gauge  the Lagrangians $L$  have the physical dimension of a  spatial energy density $[L]=EL^{-3}$, and $\mathcal{L}= L \sqrt{|g|}$ essentially that of an action $[c^{-1} \mathcal{L}]= E\, T$ (appendix  \ref{appendix dimensional considerations}). }
   
\noindent
 The Lagrange term of the scalar field   $ \mathcal{L}_{\phi}$ appears in two phases or ``regimes'', called the {\em Einstein regime}, respectively {\em Milgrom regime},  depending on the norm of the scalar field gradient (as given by the dynamical equation of the Milgrom regime),
 \beq
\mathcal{L}_{\phi} =\epsilon_{\phi} h \, \big( \mathcal{L}_{\phi_2} + \mathcal{L}_{\phi_3}+ \mathcal{L}_{_2\phi} \big) + \mathcal{L}_{V}\,  . \label{eq L-phi with h}
\eeq
The factor   $h$ is a symbolic expression for the role of a transition function $h(x, \alpha,\beta)$  which changes smoothly and monotonously  between $0$ in the Einstein regime and $1$ in the Milgrom regime, see below \eqref{eq transition function}. 
 One has to keep in mind that  we have only 
 sparse empirical information and nearly no theoretical  knowledge about the transition between the Einstein/Newton and Milgrom/MOND regimes (but see section \ref{subsection Milky Way}). For the time being, any Lagrangian for it will  be  unreliable and at best metaphoric. Therefore  the dynamics  in the   transition zone between the two regimes is not  derived from the Lagrangian \eqref{eq L-phi with h}. The intermediate dynamics is rather  modelled by a {\em  smooth transition, expressed by  $h$,  between  field solutions} formally calculated in the  two adjacent regimes.  

 The factor $\epsilon_{\phi}$  suppresses  the subsequent Lagrange terms for $\phi$ whenever the gradient of  scalar field  becomes timelike. Introducing the sign $s_{\phi}$ of the scalar field gradient we define by
\beq
s_{\phi} = \mathrm{sig}\,D\phi :=  \mathrm{sig}\,(D_\lambda \phi D^\lambda \phi)\, , \qquad \epsilon_{\phi}= \frac{1}{2}(1+ s_{\phi})\, . \label{eq convention epsilon-phi} 
 \eeq
\noindent
In the Einstein regime and on cosmological scales the scalar field Lagrangian reduces to a  quartic potential
\[
\mathcal{L}_{\phi} =\mathcal{L}_{V}\, .
\]

In \eqref{eq Lagrangian general form} $Y$ stands for matter fields based on standard model physics (baryonic matter, elementary particles, electromagnetism). Baryonic matter  is assumed to couple to the Riemannian component $g_E$ of the  Einstein gauged  Weylian metric. Formally the matter Lagrangian may be written in scale invariant form; this is expressed above by  putting $\mathcal{L}_{bar}(g_E,Y)$  in square brackets.

$\mathcal{L}_{\phi_2}$ is an  ordinary
quadratic kinetic term, $\mathcal{L}_{\phi_3}$ a  cubic kinetic term   similar to  Bekenstein's  ``aquadratic'' Lagrangian for MOND,\footnote{The  functional expression $F(|\nabla \phi|)$ used  in \citep{Bekenstein/Milgrom:1984}  relating the Lagrangian to a MONDian  interpolation   function $\mu$  
  would  be { inconsistent with the scale invariance} of the Lagrangian.  \label{eq MOND mu nu}}  
  $\mathcal{L}_{_2\phi}$ is an additional second order derivative term which endows the scalar field with non-negligible mass-energy without raising the order of the scalar field equation. $\mathcal{L}_V$ denotes the well known  quartic potential term of $\phi$.   A  timelike unit  vector field (with regard to $g$)  $A^{\mu}$ of weight $w(A)=-1$  has to be  assumed as an additional non-dynamical structure for the mass term.\footnote{$A^{\mu}$ is important for  the weak field  approximation of the scalar field equation, see below. }

In scale invariant form the contributions to $\mathcal{L}_{\phi}$ are
\begin{subequations}
\beqarr 
\mathcal{L}_{\phi_2} &=&- (\hbar c)^{-1}\frac{\alpha}{2}\,D_{\lambda}(\xi\phi)D^{\lambda}(\xi \phi) \sqrt{|g|} =  - (\hbar c)^{-1}\frac{\alpha}{2}\xi^2\,s_{\phi} |D\phi|^2\, \sqrt{|g|}      \label{eq Lagrangian phi-2}  \\
 \mathcal{L}_{\phi_3} &=&  -  \frac{\beta}{3} \phi^{-2} (s_{\phi} D_{\lambda}(\xi \phi)D^{\lambda}(\xi \phi) )^{\frac{3}{2}}\,\sqrt{|g|} =  -  \frac{\beta}{3} \xi^3 \phi^{-2}|D\phi |^3\,\sqrt{|g|}   \label{eq Lagrangian phi-3}  \\ 
 \mathcal{L}_{_2\phi} &=& (\hbar c)^{-1}\, (\xi \phi)\,D_{\lambda}D^{\lambda}(\xi \phi) A_{\lambda}A^{\lambda}\,  \sqrt{|g|}   \label{eq Lagrangian 2-phi}  \\
 \mathcal{L}_{V} &=& - (\hbar c)^{-3} V(\phi)  \,\sqrt{|g|} \, , \qquad \mbox{here} \quad V(\phi)=\lambda\, \phi^4 \, .  
 \eeqarr
 \end{subequations}
   $\xi$  is a hierarchy factor mediating between the energy levels of the Hilbert term (Planck energy $E_P$) and the cosmologically small energy $E_M=a_0\hbar$  (see  below).
Later  we will find reasons to set $\alpha = -4,\, \beta= 2$    in  the {\em relativistic MOND/Milgrom regime}, while effectively $\alpha=\beta = 0$   in the {\em Einstein/Newton} regime.

Let us  call the  (constant) value of $\phi$ in the Einstein gauge  $\phi_0$. If we write the scalar field in the  Riemann gauge in exponential form,
\beq \phi(x) \underset{Eg}{\doteq} \phi_0 \, e^{-\sigma(x)} \, , \label{eq definition sigma}
\eeq
 the scale connection in the Einstein gauged Weylian metric becomes
\[  \varphi \underset{Eg}{\doteq} d\sigma \, .
\]
Here (and elsewhere) $\underset{Eg}{\doteq}$ is used to denote equality in the Einstein gauge (see appendix \ref{appendix Weyl geometry}).

The  {\em Lagrangians} in  {\em Einstein gauge} (with identities in this gauge denoted by $ \underset{Eg}{\doteq} $) turn into:
\begin{subequations}
\beqarr  \mathcal{L}_H &\underset{Eg}{\doteq}& \frac{(\hbar c)^{-1}}{2} (\xi \phi_0)^2 R(g,d\sigma) \, \sqrt{|g|} = \frac{(8 \pi \varkappa)^{-1}}{2}\, R(g,d\sigma) \, \sqrt{|g|} \qquad (g=g_E)  \\
\mathcal{L}_{\phi_2} &\underset{Eg}{\doteq}& -(\hbar c)^{-1} \frac{\alpha}{2} (\xi \phi_0)^2 \partial_{\lambda} \sigma\, \partial^{\lambda}\sigma\,\sqrt{|g|} = - (8 \pi \varkappa)^{-1}\, \frac{\alpha}{2} (\xi \phi_0)^2 \partial_{\lambda} \sigma\, \partial^{\lambda}\sigma\,\sqrt{|g|} \label{eq Lagrangian} \\
 \mathcal{L}_{\phi_3} &\underset{Eg}{\doteq}&  =  -  \frac{\beta}{3}  (\xi^{-1}\phi_0)^{-1} (\xi \phi_0)^2  |\nabla \sigma |^3\,\sqrt{|g|} = -  \frac{\beta}{3} a_1^{-1} (8 \pi \varkappa)^{-1}(s_{\phi}\, \partial_{\lambda} \sigma\,\partial^{\lambda} \sigma )^{\frac{3}{2}}\,\sqrt{|g|}  \\
 \mathcal{L}_{_2\phi}  &\underset{Eg}{\doteq}&  -  (8 \pi \varkappa)^{-1}\, \big(\nabla_{\lambda}\partial^{\lambda}\sigma + \partial_{\lambda}\sigma \partial^{\lambda}\sigma\big) A_{\lambda}A^{\lambda}\, \sqrt{|g|}\,     \\  
 \mathcal{L}_{V} &\underset{Eg}{\doteq}& - (\hbar c)^{-3} \lambda\,(\xi^{-1}\phi_0)^2 (\xi \phi_0)^2 \,\sqrt{|g|} \underset{Eg}{\doteq}   -  \lambda a_0^2\,c^{-2} (8 \pi \varkappa)^{-1} \sqrt{|g|} \underset{Eg}{\doteq} -\frac{\Lambda}{8 \pi \varkappa} \sqrt{|g|}
 \,   \quad  \\
  \mathcal{L}_{m} &\underset{Eg}{\doteq}& \ldots \quad \mbox{(dependent on context)} 
 \eeqarr 
\end{subequations}
With constants $\xi, \varkappa$ defined   in  \eqref{eq xi und phi0} and the line above it.
The Lagrangian functions $L= \mathfrak{L}/ \sqrt{|g|}$ have the dimension  of an energy density, $[L]= E\,L^{-3}$,  the Lagrange densities $ \mathfrak{L}$ essentially that of an action, more precisely $[c^{-1}\mathcal{L}]=E\,T$ (appendix \ref{appendix dimensional considerations}).

The transition between the regimes is  modelled by a standard function  used in  differential topology for establishing ($C^{\infty}$-) smooth  transitions between two structures:
\beqarr
g(x) &=& \frac{ f(x)}{f(x)+  f(1-x)}\, , \qquad  f(x) = 
 \Big\{
{ e^{-\frac{1}{x}} \quad \mbox{for}\quad x>0   
\atop \; \; 0 \qquad \;   \mbox{for} \quad x \leq 0\, , } \nonumber \\
 h(x;{\alpha},{\beta}) &=& 1-   g\big(\frac{x-{\alpha}}{{\beta}-{\alpha}}\big)\,,\;\;\;\; \; \label{eq transition function}
\eeqarr  
where ${\alpha},\, {\beta}$ define the boundaries of the  transition regime. 
The variable $x$ may stand for the ratio of the gradient of the scalar field  to the MOND constant, $ x = \frac{|\nabla \sigma|}{a_1} $.\footnote{In the weak field approximation $\sigma$ is related to the Newton potential of baryonic matter by the equations \eqref{eq algebraic transformation a} \eqref{eq definition sigma-bar}; the threshold condition can then also be given in terms of accelerations $|a_N|$ like in classical MOND; i.e. for the variable $y=\frac{|a_N|}{\mathfrak{a}_0} $ (with modified boundaries $\tilde{\alpha}, \, \tilde{\beta}$ of the transition interval). For a centrally symmetric constellation, also the coordinate distance $r$ from the center can take the role of $x$, with a corresponding adaptation of the boundaries of the transition interval (like in section 
 \ref{section central symmetry}).
 }

\begin{figure}[h]
\centerline{ \includegraphics[scale=0.7]{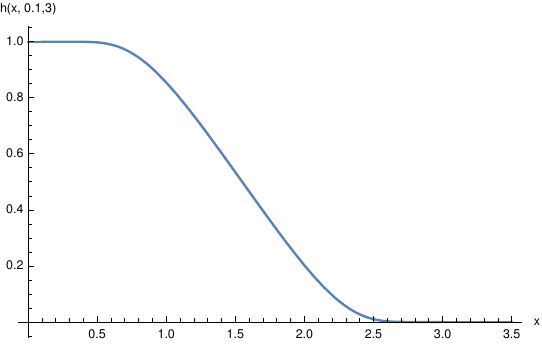} }
\caption{\small Transition function $h(x;{\alpha},{\beta})$ for   ${\alpha}=0.1,\,{\beta}=3$.    \label{fig transition function}}
\end{figure}

The suppression of the Lagrange terms \eqref{eq Lagrangian phi-2},  \eqref{eq Lagrangian phi-3}, \eqref{eq Lagrangian 2-phi}   in the Einstein regime by the factor $h$ takes up  a proposal by Berezhiani and Khoury,  developed in their superfluid approach to  dark matter and modified gravity  in \citep{Berezhiani/Khoury:2015,Berezhiani/Khoury:2016}.  
Like in  this approach we assume  that the scalar field 
can  reside in the  typical state of the   relativistic Milgrom regime only if the norm   $| \nabla \sigma |$  in the Einstein gauge   remains below a certain threshold. Once the gradient  of $ \sigma$ 
surpasses the threshold the scalar field destabilizes and 
a transition phase is entered.  Let us call this the {\em lower transition zone}.\footnote{In the following, \eqref{eq transition parameters model galaxy}, we assume $(\alpha,\beta)\approx (0.32, \,1.75)$ and $(\bar{\alpha}, \bar{\beta})\approx (0.1, \, 3)$. \label{fn alpha beta}} 

 For an even larger value of the gradient (as it would be according to the dynamical equations of the Milgrom regime) the dynamics of the actual scalar field $\bar{\sigma}$ is completely suppressed; $\nabla \bar{\sigma} \rightarrow 0$. 
  With $\bar{\sigma}= const$ the scalar field in the Riemann gauge becomes constant,
 \[
 \phi \underset{Rg}{\doteq} \phi_0 \, .
 \]
Because of the factor $\epsilon_{\phi}$ in \eqref{eq L-phi with h} the same holds for regions in which the model  would indicate a shift of the gradient from  spacelike  to timelike sign. 

 Finally a  suppression of the actual scalar field  has to be implemented  for a region in which the applicability of the model fades out, called {\em uppper transition zone},  by the complementary smooth transition function 
\[
\tilde{h}(x,\hat{\alpha},\hat{\beta})= 1- g(\frac{x-\hat{\alpha}}{\hat{\beta}-\hat{\alpha}}) \, .
\]

In the following  the usual notation
  \[
\varkappa = G\, c^{-4}\,   \qquad \mbox{($G$ the Newton gravitational constant)} 
\]
is used. Moreover we   choose  
   $\phi_0$ and the hierarchy factor $\xi$  such that in the Einstein gauge  $\xi\phi_0$ is the Planck energy and $\xi^{-1}\phi_0$ is the energy associated to the MOND constant $a_0$: 
\beq
(\xi \phi_0) = E_{P}=\big((8 \pi G)^{-1}\, \hbar c^5\big)^{\frac{1}{2}} = \big((8\pi\,\varkappa)^{-1}\, \hbar c \big)^{\frac{1}{2}}\qquad \mbox{and} \quad  \xi^{-1}\phi_0 = a_0\, \hbar \,  \label{eq xi und phi0}
\eeq
This means
\[
 (\xi \phi_0)^{-2} \hbar c = 8 \pi \varkappa.
\]

\noindent 
  $a_0$  is a   constant of physical dimension $[a_0]=T^{-1}$ with  companion $a_1=a_0\, c^{-1}$ (dimension $[a_1]=L^{-1}$),\footnote{Here and at many other places square brackets are used to denote the physical dimension of a quantity (see appendix \ref{appendix dimensional considerations}). }  
  related to the  MOND acceleration  $\mathfrak{a}_0$
 by  $ \mathfrak{a}_0= a_0\, c$.  
 The  empirical value  of this acceleration 
 \[
 \mathfrak{a}_0 \approx 1.8\cdot10^{-8}\, cm\, s^{-2} \, ,
 \]
results in the known approximation
\[ 
a_0 \approx \frac{H_0}{6} \quad \longleftrightarrow \quad  a_1 = a_0\, c^{-1}\approx \frac{H_1}{6} \qquad \qquad (H_1 = H_0 c^{-1})\,.
  \]
\noindent
    The hierarchy factor $\xi$ is  chosen such that $\phi_0$ is {\em placed at  the geometrical mean}  between  the cosmologically small   energy scale $E_M= a_0\hbar \sim 10^{-32}\, eV $  (  $\sim \frac{H_0}{6} \hbar $) which may be called {\em  Milgrom energy}  and the reduced Planck energy $E_P\sim 10^{27}\, eV$.  Its order of magnitude is $\xi \sim  10^{30}$. Of course, this  convention  may be changed if preferred.  
    
In the  Einstein gauge the quartic potential $L_V$ degenerates into a cosmological constant term with coefficient $\frac{\Lambda}{8\pi G}c^{-2}$. The above choice of $\phi_0$ (in the geometric mean of the extremes) results in a moderate   order of magnitude for  $\lambda$.  Using  the usual notation $\Omega_{\Lambda}$ one finds:
\beq
\lambda\,  a_1^2 =  \Lambda = 3 \, \Omega_{\lambda}\, H_1^2 \qquad \longleftrightarrow \quad \lambda \approx 3\cdot 6^2 \, \Omega_{\Lambda}\, 
 \eeq

\subsection{\small Dynamical equations \label{subsection dynamical equations}}

\noindent
\textbf{Einstein equation}\\
The  variational derivation $\delta g^{\mu\nu}$  of the   Hilbert term \eqref{eq Hilbert Lagrangian} leads to  the  Weyl geometric Einstein tensor $G_W = G(g, d\varphi)$ plus an additional term    due to the variation of the non-minimally coupled scalar field:\footnote{See \citep[eq. (2.17)]{Drechsler/Tann}; a similar term is known in JBD theory \citep[eqs. (3.1), (3.5)]{Capozziello/Faraoni}, \citep[(2.1), (2.6))]{Fujii/Maeda}. 
} 
\[
(\xi \phi)^{-2} \big(g_{\mu\nu}\, D_{\lambda}D^{\lambda}(\xi \phi)^2 - D_{(\mu}D_{\nu)}(\xi\phi)^2  \big) \, 
\]
Brought to the right hand side (r.h.s.) as  
\[
\Theta^{(\phi_H)} =  - (\xi \phi)^{-2} \big(g_{\mu\nu}\, D_{\lambda}D^{\lambda}(\xi \phi)^2 - D_{(\mu}D_{\nu)}(\xi\phi)^2  \big)
\]
it contributes to the effective scalar field energy momentum  of the vacuum Einstein equation 
\[
 G_W = G(g, d\varphi) =  \Theta^{(\phi)} +  \Theta^{(\phi_H)}
 \, .  
\]
 $ \Theta^{(\phi)} $ is the energy term arising from the variational derivative  of $\mathcal{L}_{\phi}$ (here denoted by  $[ \mathcal{L}_{\phi} ]_{g^{\mu\nu}}$),
\[
\Theta^{(\phi)} =  - \frac{2 (\hbar c)}{\sqrt{|g|}} (\xi \phi)^{-2} [ \mathcal{L}_{\phi} ]_{g^{\mu\nu}} \underset{Eg}{\doteq} -\frac{2}{\sqrt{|g|}}\, 8\pi \varkappa \,[\mathcal{L}_{\phi}]_{g^{\mu\nu}}\, .
\]
Its corresponding energy tensor is 
\[
T^{(\phi)}_{\mu\nu} = - \frac{2}{\sqrt{|g|}} {[}\mathcal{L}_{\phi} {]}_{g^{\mu\nu}} = (\hbar c)(\xi \phi)^2 \Theta^{(\phi)} \underset{Eg}{\doteq} 8 \pi \varkappa\, \Theta^{(\phi)} \, .
\]

If  baryonic matter is included a similar energy expression for baryonic matter $\Theta^{(bar)}$  appears on the r.h.s. of the Einstein equation, corresponding to a matter  energy-momentum tensor formally written in scale covariant form,
\[
T^{(bar)}_{\mu \nu} =  - \frac{2}{\sqrt{|g|}} {[}\mathcal{L}_{bar} {]}_{g^{\mu\nu}}  = (\hbar c)(\xi \phi)^2 \Theta^{(bar)} \, .
\]

Because of the coupling condition for matter the Einstein equation with baryonic matter is best expressed in the {\em Einstein gauge} 
\beq G_W \underset{Eg}{\doteq} G(g_E, d\sigma) \underset{Eg}{\doteq} (8\pi\varkappa)\, T^{(bar)}_E + \Theta^{(\phi)}_E + \Theta^{(\phi_H)}_E  \, .   \label{eq Einstein eq with matter}
\eeq 
The subscript $E$ indicates that all terms on the r.h.s. are understood in their Einstein gauge. In the following this will be presupposed  even without the subscript for the r.h.s of a $\underset{Eg}{\doteq} $ relation. 
For bringing the form of \eqref{eq Einstein eq with matter} closer  to Einstein gravity we decompose the l.h.s. into its Riemannian component  and the scale connection contribution (see appendix \ref{appendix Weyl geometry})
\[
G_W   \underset{Eg}{\doteq} G(g) + G(d\sigma) \, ,
\]
with $G(d\sigma)$ give by \eqref{eq sigma contribution Einstein tensor}.
Shifting it to the r.h.s becomes
\beq 
G(g)    \underset{Eg}{\doteq} (8\pi\varkappa)\, T^{(bar)} + \Theta^{(\phi)} + \Theta^{(\phi_H)}  - G(d\sigma) \, . \label{Weylgeometric EEq 0}
\eeq
 
In dimension $n=4$ the second order derivative terms of $\Theta^{(\phi_H)}$ and $G(d\sigma)$ cancel, 
the  added contributions of the scalar field  in \eqref{Weylgeometric EEq 0}
simplify to:  
 \beq
 \bar{\Theta}^{(\phi)}_{\mu\nu} \underset{Eg}{\doteq} \Theta^{(\phi)}_{\mu\nu} -   3 \partial_{\lambda}\sigma \partial^{\lambda}\sigma\, g_{\mu \nu} + 6 \partial_{\mu}\sigma \partial_{\nu}\sigma  \,.  \label{eq full scalar field energy expression}
 \eeq
$\bar{\Theta}^{(\phi)}$ sums up all contributions  of  the scalar field to the r.h.s. of the Einstein equation, including its  gravitational contributions (deriving from the Hilbert term and the scale connection part of the Einstein tensor) on a par with  its  energy expression proper $\Theta^{(\phi)}$.
The final form of the gravitational equation for $n=4$ in the Einstein gauge is then
\beq 
G(g)    \underset{Eg}{\doteq} (8\pi\varkappa)\, T^{(bar)}_E + \bar{\Theta}^{(\phi)}\, . \label{Weylgeometric EEq}
\eeq
Note, however, that neither $ \bar{\Theta}^{(\phi)}$ nor its  energy tensor 
\[
\bar{T}^{(\phi)}_{\mu\nu}  \underset{Eg}{\doteq} (8 \pi \varkappa)^{-1} \bar{\Theta}^{(\phi)}_{\mu \nu} \, 
\]
 is  scale covariant because of the contributions from $G(\varphi)$. It  will thus  be used in the Einstein gauge only.
 
 After evaluating all contributions to $\Theta^{(\phi)}$  in the Milgrom regime  
  this is 
\beqarr
\bar{\Theta}^{(\phi)}_{\mu \nu}  &\underset{Eg}{\doteq}&  
\Box(g)\sigma(g_{\mu\nu} + 2 A_{\mu}A_{\nu}) - 2 \nabla(g)_{(\mu}\partial_{\nu)} \sigma    \label{eq Theta-phi-bar} \\
  & & + (\alpha+12)\,\partial_{\mu}\sigma\partial_{\nu}\sigma + 2 \,\partial_{\lambda}\sigma\partial^{\lambda}\sigma A_{\mu}A_{\nu} - \big((\frac{\alpha}{2} +3)\partial_{\lambda}\sigma\partial^{\lambda}\sigma   +  \frac{2 \beta}{3} a_1^{-1}\, | \nabla
 \sigma |^3 + \Lambda \big)g_{\mu\nu}\, .   \nonumber
\eeqarr
Outside the Milgrom regime it reduces to  $\bar{\Theta}^{(\phi)}_{\mu \nu}  \underset{Eg}{\doteq} \Lambda g_{\mu\nu}$, i.e., the cosmological term of  Einstein gravity.  \\[0.5em]

\noindent
\textbf{Scalar field equation}\\
Baryonic matter does not couple directly to the scalar field.  In the Milgrom regime the scalar field equation 
\[
[\mathcal{L}]_{\phi} =  [{L}]_{\phi} = \sum_{X\in {\phi_2,\phi_3,_2\phi,V}}  [{L}_X]_{\phi}  = 0 
\]
gives term by term:\footnote{Cf. appendix \ref{appendix variations}.}
\beqarr
 2 \phi^{-1} {L}_H  
&+&  \alpha \xi^2 D_{\lambda} D^{\lambda}\phi   \label{eq rough scalar field equation} \\
&+&  4 \phi^{-1}  {L}_{\phi_3} +   \,s_{\phi}\,\, \beta\, \xi^3  \phi^{-2} D_{\lambda}\big(|D \phi|D^{\lambda}\phi \big) 
+ 2 \, \xi^2 D_{\hspace{-0.1em}\lambda}D^{\lambda} \phi\,A_{\lambda}A^{\lambda} 
+  4 \phi^{-1} {L}_{V}  = 0 \nonumber
\eeqarr
On shell of the Einstein equation  
 the scalar field equation 
is obviously equivalent  to 
\[
2\, tr\,[\mathcal{L}]_g + \phi [\mathcal{L}]_{\phi} = 0 \, . 
\]
This condition for $\phi$  will be called   the  {\em reduced scalar field equation}.  Like in JBD theory it encodes an indirect coupling,  mediated by the Hilbert term, between baryonic matter and the scalar field. 

\noindent
With \eqref{eq trace L-g}, \eqref{eq D-lambda-square phi-square},  \eqref{eq rough scalar field equation} and setting
 \beq
 \alpha=-4 \label{eq alpha}
 \eeq
 the reduced scalar field equation becomes
\beqa
 s_{\phi} \beta\,\xi^3 \phi^{-1}\, D_{\lambda} (|D\phi|  D^{\lambda}\phi)  &+&      2 \,(\hbar c)^{-1}\,\xi^2 D_{\lambda}\phi D^{\lambda}\phi   + 3 \mathcal{L}_{\phi_3} \nonumber \\
  &+&   2\,(\hbar c)^{-1}\,\xi^2 (1 +  A_{\lambda}A^{\lambda} )\,  \phi\, D_{\lambda}D^{\lambda}\phi =  tr\, [T^{(bar)}]\, . 
\eeqa
The unit condition for the non-dynamical vector field $A^{\mu}$  simplifies it to
\beq
 s_{\phi} \beta\,\xi^3 \phi^{-1}\, D_{\lambda} (|D\phi|  D^{\lambda}\phi)  +      2 \,(\hbar c)^{-1}\,\xi^2 D_{\lambda}\phi D^{\lambda}\phi   + 3 \mathcal{L}_{\phi_3} 
  =  tr\, [T^{(bar)}]\, . \label{eq reduced SF long} 
\eeq
\noindent
 For    spacelike $\nabla \sigma$  in the Milgrom regime   ($ s_{\phi}=1$)
this   is\footnote{Note that \eqref{eq xi und phi0} implies 
$(\xi^{-1}\phi_0)^{-1}(\xi \phi_0)^2 =(a_0\hbar)^{-1} (8\pi \varkappa)^{-1}\, \hbar c = (a_0\,c^{-1})^{-1}\, (8\pi \varkappa)^{-1} = a_1^{-1}(8 \pi \varkappa)^{-1}    \, .
$
 }
 \beq  \nabla(g)_{\lambda}(|\nabla \sigma |\partial^{\lambda}\sigma \big) 
- 2 \beta^{-1}\, a_1 |\nabla \sigma|^2  + |\nabla \sigma|^3 
 \underset{Eg}{\doteq} - a_1 \, \beta^{-1} (8 \pi \varkappa)\, tr\, T^{(bar)}   \, .  
 \label{eq full Milgrom equation}
\eeq
 Both sides of the equation are of dim $L^{-3}$.

The  last two terms on the l.h.s. together form a cubic polynomial in $|\nabla \sigma|$, 
\[
p_3(x) = (x - 2 \beta^{-1} a_1)\, x^2 \, .
\] 
In  the Milgrom regime  with  $|\nabla\sigma| \leq a_1$  and $\beta = 2$ it is negligible: 
   \[
    |p_3(|\sigma|)| \leq   c^{-3} a_0^3 < c^{-3}H_0^3 \ll \Lambda \, 
    \]
Therefore it is justified to represent  \eqref{eq full Milgrom equation}  in the simpler approximate form 
\beq
 \nabla(g)_{\lambda}(|\nabla \sigma |\partial^{\lambda}\sigma \big) \underset{Eg}{\doteq} - \mathfrak{a}_0 \beta^{-1}\, (8 \pi \varkappa)\, c^{-2}\,tr\, T^{(bar)}   \, . \label{eq core Milgrom equation}
\eeq
This approximation will be called the     {\em relativistic Milgrom equation} of the present model. It is a straight forward covariant generalization of  the nonlinear Poisson equation of deep  MOND. 

\subsection{\small The scalar field energy tensor in weak fields \label{subsection weak fields}}
The first order derivatives appear in quadratic or cubic monomials. In  weak field constellations
the energy-momentum  is therefore dominated by the second order derivative terms of $\sigma$, because .
We thus concentrate on the    second order expressions  in \eqref{eq Theta-phi-bar}  denoted by $\hat{\Theta}^{(\phi)}$ and  $\hat{T}^{(\phi)}$:
 \begin{subequations}
 \beqarr
\hat{\Theta}^{(\phi)}_{\mu\nu} &\underset{Eg}{\doteq}&  \Big( \Box(g)\sigma \, (g_{\mu\nu} + 2\,  A_{\mu}A_{\nu}) - 2\, \nabla(g)_{(\mu} \partial_{\nu)}\sigma \Big) \,   \label{eq Theta-bar} \\
 \hat{T}^{(\phi)}_{\mu\nu} &\underset{Eg}{\doteq}& (8 \pi \varkappa)^{-1} \hat{\Theta}^{(\phi)}_{\mu\nu}  \,  \label{eq T-phi-bar}
\eeqarr
 \end{subequations}
 Both are traceless.\\
\eqref{eq Theta-phi-bar}  becomes then
\[
\bar{\Theta}_{\mu\nu} \underset{Eg}=  \hat{\Theta}_{\mu\nu} + p_{\mu\nu} + \Lambda g_{\mu\nu}\, ,
\]
where  $p_{\mu\nu}$ are the components of a tensor $p(d\sigma)$ given by third degree polynomials  in the partial derivatives.\footnote{
$
p_{\mu\nu} =  (\alpha+12)\,\partial_{\mu}\sigma\partial_{\nu}\sigma + 2 \, \partial_{\lambda}\sigma\partial^{\lambda}\sigma A_{\mu}A_{\nu} - \big((\frac{\alpha}{2} +3) \partial_{\lambda}\sigma\partial^{\lambda}\sigma   + \epsilon_{\phi}\frac{2\beta}{3} a_1^{-1}\, | \nabla
 \sigma |^3  \big)g_{\mu\nu}\,  
$
}

 \noindent
 Focussing  attention on the second order terms the full the energy tensor can be written as
\beq
\bar{T}^{(\phi)} \underset{Eg}=  \hat{T}^{(\phi)} +  (8\pi \varkappa)^{-1}\big(  \Lambda\, g_{\mu\nu} + p(\partial\sigma) \big)\,  \label{eq full energy tensor phi}
\eeq
and the Einstein equation as 
\beq
G(g) \underset{Eg}=   8\pi \varkappa\, \big(T^{(bar)} + \hat{T}^{(\phi)} \big) + \Lambda g +  p(\partial\sigma) \, .
\label{eq Einstein equation}
\eeq 
 In cases where  the  polynomial and the cosmological term can be neglected, the full Einstein equation may be substituted by the approximation
\beq
G(g) \underset{Eg} \approx   8\pi \varkappa\, \big(T^{(bar)} + \hat{T}^{(\phi)}\big) \, ; \label{eq Einstein equ simplified}
\eeq
remember that $\hat{T}^{(\phi)}$ contains only second order derivates  of $\sigma$.
\noindent

For quasi-static spacelike  $\nabla \sigma$,   $A_0=1$ and $g_{00}\approx -1$ the dominant second order terms of the energy component of  \eqref{eq T-phi-bar} reduce to\footnote{For dimensional considerations the dimension $L^2$ of the hidden factor $g_{00}$ has to be plugged in, such that $[\hat{T}^{(\phi)}_{00} ]= EL^{-1}L^{-2}L^2= EL^{-1}$).}
\beq
\rho^{(\phi)} \underset{Eg} = \hat{T}^{(\phi)}_{00}  \approx (8 \pi \varkappa)^{-1}  \,  \Box(g)\sigma =   (8 \pi \varkappa)^{-1}  \, \nabla^2\hspace{-0.1em}(g) \sigma \,  , 
\,  \label{eq energy component scalar field}
\eeq
with $ \nabla^2\hspace{-0.1em}(g)$ the Laplacian with regard to $g$.

\noindent
Because of $tr\, \hat{T}^{(\phi)}= 0$ the expression entering the weak field approximation  of Einstein gravity is also
\beq
\hat{T}^{(\phi)}_{00} - \frac{1}{2}tr\, \hat{T}^{(\phi)}g_{00} \approx  (8 \pi \varkappa)^{-1}  \, \nabla^2\hspace{-0.1em}(g)\sigma \,  . \label{eq rhs scalar field contribution to Newton approximation}
\eeq

\section{\small Flat space approximation: a special type of MOND dynamics \label{section flat approximation}}

\subsection{\small Newton approximation and deep MOND equation for the scalar field \label{subsection flat space limit}}
\textbf{Gravity}\\
Let us consider a Riemannian component of the metric for a  quasi-static mass distribution, 
\beq
g_{\mu\nu} = \eta_{\mu\nu} + h_{\mu  \nu}  \qquad\mbox{(signature (-+++))}\, , \label{eq weak field metric}
\eeq  
where  $\eta$ is the Minkowski metric and $h$ is diagonal with small entries (both in special relativistic dimensional conventions $[\eta]=[h]=1$). For the  Einstein equation rewritten as
\[
Ric = 8 \pi \varkappa \big(T- \frac{1}{2}tr\, T\, g  \big) 
\]
there is  a well known first order approximation of the energy component of the l.h.s.:
\beq
R_{00} \approx - \frac{1}{2} \eta^{\lambda \lambda}\partial_{\lambda} \partial_{\lambda}h_{00}= - \frac{1}{2} \delta^{jj} \partial_j\partial_j h_{00} \, .  \label{eq Ri0j0 weak field} 
\eeq
After defining a potential
\beq
\Phi := -\frac{1}{2}c^2\, h_{00} \label{eq Phi -- h00}
\eeq
the ensuing relation
\beq
\nabla^2 \Phi \approx c^2 R_{00} = (8\pi \varkappa) c^{2}(T_{00}-\frac{1}{2}tr T\, g_{00}) = (4\pi G)c^{-2}\, (2\, T_{00}-tr T\, g_{00})\,  \label{eq Poisson equation Newton approximation 0}
\eeq
is the Newton approximation of Einstein gravity.  For $T^{(bar)} $ of pressure free matter with energy density $\rho_e$, i.e.,  mass density $\rho_m= c^{-2} \rho_e$, this is the usual Poisson equation $\nabla^2 \Phi = 4\pi G\, \rho_m$.

In our case with $T=T^{(bar)} + \hat{T}^{(\phi)}$ we  add the scalar field contribution using  \eqref{eq rhs scalar field contribution to Newton approximation}
\[ 
(8\pi G)c^{-2}\, (\hat{T}^{(\phi)}_{00} - \frac{1}{2} tr \hat{T}^{(\phi)} g_{00} ) \approx  \nabla^2\hspace{-0.1em}(g)  \sigma\, .
\]
According to \eqref{eq energy component scalar field} this is 
\[
 \nabla^2\hspace{-0.1em}(g)  \sigma = (4 \pi G)\,c^{-2}\,  2\, \rho^{(\phi)} = 4\pi G\, 2\, \rho^{(\phi)}_m \, .
\]
The symbol $\rho^{(\phi)}_m $ is  here used for  
the {\em mass density}  corresponding to the scalar field energy density,
\beq \rho^{(\phi)}_m = c^{-2} \rho^{(\phi)} \, , \label{eq mass density scalar field}
\eeq
 {\em without} raising an ontological  claim of the ``mass''-ness for the scalar field. It differs from the Newtonian mass equivalent of the scalar field energy, which one would expect from the  Poisson equation of pressure free matter (see \eqref{eq Phi-phi}).
\\[-0.2em]

Evaluation of  \eqref{eq Poisson equation Newton approximation 0}
for pressure free baryonic matter and the scalar field together  results in a Poisson equation of the form
\beq
\nabla^2 \Phi = 4\pi G\, \big(\rho^{(bar)}_m + 2\, \rho^{(\phi)}_m  \big)\, \label{eq Poisson equation Newton approximation}
\eeq
(denoting here the baryonic mass density by $\rho^{(bar)}_m $ and with ``$=$''  the equality in the flat space approximation).

\noindent
The total Newton potential is a superposition  of contributions from the baryonic matter and the scalar field
\beq
\Phi =  \Phi^{(bar)} +  \Phi^{(\phi)}\,  \qquad \mbox{with}\quad  \nabla^2 \Phi^{(bar)} = 4\pi G \, \rho_m\,, \quad  \nabla^2 \Phi^{(\phi)} = 4 \pi G\, 2 \rho^{(\phi)}_m\,. \label{eq total potential Newton approximation}
\eeq 
Because of the pressure term the relativistic mass-energy density of the scalar field enters  the potential of the  Newton approximation  with  (relative) factor  2. If one defines the {\em Newtonian mass equivalent} $ \rho^{(\phi)}_N $ of the scalar field 
as the source of  the usual Newtonian Poisson equation,
\[
 \Delta \Phi^{(\phi)} = 4\pi G\, \rho^{(\phi)}_N\, , 
\]
 one finds
\beq
 \rho^{(\phi)}_N = c^{-2} 2\, \rho^{(\phi)} \, . \label{eq Newtonian mass equivalent phi 2}
\eeq 
It differs  from the ``true''  mass density associated to the (field theoretic) energy density of the scalar field \eqref{eq mass density scalar field}  by the factor $2$. In certain contexts it is called the density of ``phantom matter''.

A comparison  with \eqref{eq energy component scalar field} 
 shows 
\beq
\Phi^{(\phi)} = c^2\, \sigma  \, .\label{eq Phi-phi}
\eeq
 Up to the  factor $c^2$    the scalar field potential of the flat space approximation is given by the  exponential $\sigma$ of the scalar field in Riemann gauge \eqref{eq definition sigma}.
 
The total acceleration $a$ of test masses in the  Newton approximation is 
\beq
a = -\nabla \Phi= a^{(bar)} +a^{(\phi)} \qquad \mbox{with} \quad 
a^{(bar)}= - \nabla \Phi^{(bar)}\,, \quad 
a^{(\phi)} = -\nabla \Phi^{(\phi)}= - c^2\,\nabla \sigma \, 
\label{eq additional acceleration phi}
\eeq 
and the boundaries of the transition interval of \eqref{eq transition function} may be rewritten as
\beq
 \bar{\alpha}\, a_0 \leq |a^{(\phi)}| \leq \bar{\beta}\, a_0 \qquad \mbox{with, e.g., the choice  of  } \bar{\alpha} = 0.1\,, \; \bar{\beta}=  \,3\; \mbox{(see fn.  \ref{fn alpha beta}}). \label{eq boundaries transition region}
\eeq

\noindent
\textbf{Scalar field}\\
\noindent
For pressure free matter with matter density $\rho^{(bar)}_{m}$ (dimension $ML^{-3}$) the r.h.s. of \eqref{eq core Milgrom equation} turns into $ \mathfrak{a}_0 \beta^{-1}\, 8 \pi \varkappa\, \rho^{(bar)}_m$.\footnote{For pressure  free matter with energy component of the tensor $ {T_{00}}^{(m)}=\rho^{(m)}$ ($[\rho^{(m)}]=EL^{-1}$) and energy density $\rho^{(m)}_e= - {T^{(bar)}}^0_{\;0} =-tr\, T^{(bar)} $ ($[\rho^{(m)}_e]= EL^{-3}$) the expression $- c^{-2}\,tr\, T^{(bar)}$   on the  r.h.s. reduces to the   matter density,  
$
\rho_m =  - c^{-2}\,tr\, T^{(bar)} 
$ ( with the expected dimension $[\rho^{(m)}]=ML^{-3}$);    the whole r.h.s. thus to
 $
  \mathfrak{a}_0 \beta^{-1}\, 8 \pi \varkappa\, \rho_m \, $.
  }
This  suggests to set 
\beq
 \beta = 2 \, . \label{eq beta}
\eeq

\noindent
 
By multiplying  with $c^4$  and replacing  $\sigma$  with the  potential $\Phi^{(\phi)}$,   the relativistic Milgrom equation \eqref{eq core Milgrom equation}   for pressure free quasi-static matter  turns into 
\[
 \nabla(g)_{\lambda}\,(|\nabla \Phi^{(\phi)} |\partial^{\lambda}\Phi^{(\phi)}\big)  =    
   \mathfrak{a}_0 \,  (4 \pi G)\, \rho^{(bar)}_m  \, .  \label{eq approximate Milgrom equation}
\]

   In the notation of vector calculus in the weak field limit  (with
  $\nabla(g) \rightarrow \nabla$ of flat space)  this is
\beq
 \nabla \cdot\big(|\nabla \Phi^{(\phi)}| \nabla \Phi^{(\phi)}\big)  =   \mathfrak{a}_0 \,  (4 \pi G)\, \rho^{(bar)}_m \,  .  \label{eq flat space Milgrom equation} 
\eeq
In other words, the scalar field potential $\Phi^{(\phi)}= c^2 \sigma $  \eqref{eq Phi-phi}  satisfies   the   deep MOND equation of the usual MOND appraoch.\footnote{The deep MOND equation is the non-linear Poisson equation of MOND  \citep[eq. (17)]{Famaey/McGaugh:MOND} for the interpolation function  $\mu=id$.}  \\[-0.5em]

\noindent
{\em Intermediate Result}:\\ 
The total potential in the Newton approximation \eqref{eq total potential Newton approximation} of our model is a sum of the Newtonian matter potential $\Phi^{(bar)}$ and  a  deep MOND potential $\Phi^{(\phi)}$  due to  the scalar field  \eqref{eq flat space Milgrom equation}. Together they form the {\em Newton-Milgrom  approximation} of the  scalar field model. Its phenomenology places it in the family of MOND models while it has peculiar features which give it its own distinctive flavour. They  are explored in the sequel.\\[-0.3em]

\subsection{\small Inferences from the flat space approximation on the scalar field halo 
\label{reverse Milgrom approximation}}
The calculations in the flat space approximation can be used for investigating properties of the relativistic model by  
 a kind of {\em reverse Milgrom approximation}.  
Starting from data and calculations in the flat space approximation  one can  infer the approximate value of the  scalar field potential:\\[-0.5em]
 
Assume that for a given classical baryonic mass density $\rho^{(bar)}$ in the Milgrom regime the  Newton potential $\Phi^{(bar)}$  in flat space  
is given, and its induced acceleration  is
\[ a^{(bar)} = -\nabla \Phi^{(bar)} \qquad \mbox{(dimension $LT^{-2}$)} \, .
\]
  Then the acceleration
\beq
\bar{a} =  \sqrt{\frac{\mathfrak{a}_0}{ |a^{(bar)}|}}\, a^{(bar)} =  \sqrt{\mathfrak{a}_0 |a^{(bar)}|}\, \frac{a^{(bar)}}{|a^{(bar)}|} \,  \label{eq algebraic transformation a}
\eeq  
is the (negative) gradient of a scalar potential $\bar{\Phi}$ satisfying 
   \eqref{eq flat space Milgrom equation}  like   $\Phi^{(\phi)}$.\footnote{A stepwise procedure 
 slightly more general than here, building on the equations 
 \eqref{eq algebraic transformation a}, \eqref{eq mass-energy scalar field in flat case},  is discussed in the literature  as a   ``quasi-linear'' formulation of MOND  (QMOND)  \citep{Milgrom:2010},  \citep[sec. 6.1.3]{Famaey/McGaugh:MOND}. 
Here it is not used as an alternative variant of Milgrom's dynamics but for  an inference from the flat space calculations of MOND to properties of the relativistic scalar field $\phi$. 
 \label{fn QMOND} } \\[-0.5em]

{\em Proof}: Using the duality in Euclidean space between vector fields and 1-forms,  $a^{(bar)}$ may be conceived  a closed 1-form (on flat space), $d\hspace{0.07em} a^{(bar)}=0$. Thus  also $d \bar{a}=0$;  and  the integrability of $\bar{a}$  follows (flat space is simply connected). Choose  any of   its integrals, 
\beq
\bar{\Phi} = - \int \bar{a} \,  \qquad \longleftrightarrow \qquad 
\nabla  \bar{\Phi} = - \,\bar{a} . \label{eq definition sigma-bar}
\eeq
A straight forward vector calculation shows that $\bar{\Phi}$ satisfies the equation \eqref{eq flat space Milgrom equation}. 
$_\square$\\[-0.2em]

 Because of the non-linearity of equation 
  \eqref{eq flat space Milgrom equation}  this does not in general imply equality of 
$ \bar{\Phi}$  with any other solution of the equation (up to an additive constant).\footnote{According to \citep[sec. 2]{Brada/Milgrom:1995} both are equal under specific symmetry conditions, e.g., for  centrally symmetric and axisymmetric constellations.}
 But we consider and define it as the {\em main solution} of  \eqref{eq flat space Milgrom equation} and  continue to work with
\beq
\Phi^{(\phi)} = \bar{\Phi} \, .
\eeq
Then $a_{\phi}=\bar{a}$, 
and the total acceleration in the Milgrom regime is
\beq
a = a^{(bar)} + \bar{a}= a^{(bar)} \, \big(1 + \sqrt{\frac{\mathfrak{a}_0}{|a^{(bar)}|}} \big)\, . \label{eq total acceleration Milgrom regime}
\eeq

\noindent
The fictitious matter density   $\bar{\rho}$ corresponding to $\bar{\Phi}=\Phi^{(\phi)} $ calculated in flat space according to Newton gravity is
\[
\bar{\rho}= -
 (4 \pi G)^{-1}\,\nabla \cdot \,\bar{a} =  (4 \pi G)^{-1}\, \nabla^2 \bar{\Phi} \, , 
\]
This is an approximation for the Newtonian mass equivalent of the scalar field \eqref{eq Newtonian mass equivalent phi 2} (respectively the phantom matter density mentioned above).\\
\noindent
Because of 
\eqref{eq algebraic transformation a}
\[
\nabla \cdot \bar{a} =  a^{(bar)}\cdot \nabla\big(  \sqrt{\frac{\mathfrak{a}_0}{|a^{(bar)}|}} \big)   + 
\sqrt{\frac{\mathfrak{a}_0}{|a^{(bar)}|}} \cdot \nabla a^{(bar)} \, ,
\]
and thus 
\beq
\bar{\rho} =   (8\pi G)^{-1}\sqrt{\frac{\mathfrak{a}_0}{|a^{(bar)}|^3}}\, \big(a^{(bar)}  \cdot \nabla |a^{(bar)}|\big)  +  \sqrt{\frac{\mathfrak{a}_0}{|a^{(bar)}|}} \rho^{(bar)} \,. \label{eq mass-energy scalar field in flat case}
\eeq
The first term expresses a flat space  approximation of the phantom mass-energy generated in matter free regions of a changing  Newtonian acceleration field induced by baryonic matter,  the  second one adds  energy  proportional  to the baryonic matter density. 
Because of  \eqref{eq Poisson equation Newton approximation} 
the density \eqref{eq mass-energy scalar field in flat case}  calculated according to the Newtonian Poisson equation (in this sense ``phantom'')  is  twice the energy density of the scalar field converted to mass and equal to its total Newtonian mass equivalent   \eqref{eq Newtonian mass equivalent phi 2},
\[
 \bar{\rho}   \approx \rho_N^{(\phi)}  \approx 2\, c^{-2}\rho^{(\phi)}_e \,.
\]

\subsection{\small MOND ``interpolation'' functions \label{subsection MOND interpolation functions}}
  In this subsection  $a_N$ denotes the Euclidean norm of the  Newtonian acceleration  of baryonic mass, $a_N = |a^{(bar)}|$;  the symbol $a$ is used here as an abbreviation for  the modulus of the dynamically effective total acceleration (not the whole vector as at other places). \\
In  classical (flat space) MOND  the functional relationship between $a_N $ and $a$ is expressed by two functions called ``interpolation'' functions,\footnote{The transformation functions $\mu$, $\nu$ are often considered as  ``interpolating'' between the Newtonian and deep MOND dynamics. }
\[
a = \tilde{\nu}(a_N)\, ,   \quad  \quad a_N= \tilde{\mu}(a)\,,   \quad  \mbox{with mutually inverse functions} \quad \tilde{\nu}\circ\tilde{\mu}= id\, .
\] 
 This relationship is independent of the  baryonic mass distribution of  individual galaxies. The $\tilde{\nu}$-relation plays an important role in  empirical studies;  its specification for disk galaxies  is  called the {\em radial acceleration relation} (see section \ref{subsection radial acc relation}).  

 For emphasizing the role of the typical constant acceleration $\mathfrak{a}_0= c\, a_0$ below which $\tilde{\nu}$ and $\tilde{\mu}$ deviate from the identity the transformation is usually  expressed  in the   form 
\beq \tilde{\nu}(a_N)= \nu(\frac{a_N}{\mathfrak{a}_0})a_N = a\, , \quad \tilde{\mu}(a)=\mu(\frac{a}{\mathfrak{a}_0})a = a_N \, . \label{eq mu nu MOND defined}
\eeq
In the MOND literature 
$\mu$ and $\nu$ come always in pairs but are not uniquely determined. Because of \eqref{eq mu nu MOND defined}  every pair  satisfies  the  constitutive relationship: 
\beq \mu(x)\, \nu(y)=1 \quad 
\mbox{for variables $x, \, y$ such that}   \quad \mu(x)\, x = y \; \mbox{and} \;   \nu(y) \, y = x \, . \label{constitutive relation mu nu}
\eeq
Moreover two asymptotic conditions have to be satisfied:
\beqarr
 \mu(x) \overset{x\to \infty}{\longrightarrow} 1\,,  &\quad & \nu(y)\overset{y\to \infty}{\longrightarrow} 1  \qquad \quad \mbox{(upper asymptotic)} \label{eq asymptotic conditions}\\
 \mu(x) \overset{x\to 0}{\longrightarrow} x \
  \,  , &\quad &  \nu(y) \overset{y\to 0}{\longrightarrow} y^{-\frac{1}{2}}   \qquad \mbox{(lower asymptotic)} \nonumber
\eeqarr
(where $ \mu(x) \overset{x\to 0}{\longrightarrow} x $ is understood in the sense of $ \frac{x}{\mu(x)} \overset{x\to 0}{\longrightarrow} 1$).

By setting
$ x= \frac{a}{\mathfrak{a}_0}$ {and} $y=\frac{a_N}{\mathfrak{a}_0}$
any pair of functions satisfying the constitutive relationship and the asymptotic condition defines an {\em algorithmic MOND model}  for galactic dynamics. 
The upper  condition  of \eqref{eq asymptotic conditions}  then expresses the  asymptotic transition to Newton dynamics, the lower  one the transition to the so-called {\em deep MOND} regime characterized by the limit condition
\[ a=\sqrt{\mathfrak{a}_0a_N} \, .
\]
The upper condition should  be satisfied not only asymptotically, but convert to Newtonian dynamics already  for $x$ (respectively $y$) above a finite value, at least in an extremely good approximation.

A typical family of elementary algebraic functions, indexed by $n$,  is \citep[p. 52]{Famaey/McGaugh:MOND},
\beq
\mu_n(x)= x (1+x^n)^{-\frac{1}{n}} \, , \qquad \nu_n(y) = \Big(\frac{1}{2}\big(1+ (1+ 4 y^{-n} \big)^{\frac{1}{2}} \Big)^{\frac{1}{n}}\,,  \quad n= 1, 2, 3 \ldots \label{eq simple interpolation functions}
\eeq
For basic considerations often the pair of {\em simple} transformation functions $(\mu_1, \, \nu_1)$ suffices. 
In his first study   of the so-called  ``mass deficiency -- acceleration relation''  \citep{McGaugh:2004}  McGaugh prefers the quadratic one $(\mu_2, \nu_2)$ (for a more refined study see section \ref{subsection radial acc relation}).

 We  directly read off the  MONDian interpolation function $\nu$ of our model  in the Milgrom regime from  \eqref{eq total acceleration Milgrom regime}: 
\beq
 \nu(y) = 1 + y^{-\frac{1}{2}}\,  \label{eq nu-M}
\eeq
The transition between the  Milgrom and the Einstein/Newton regimes takes place for $y=\frac{a_N}{a_0}$ between $\bar{\alpha}$ and $\bar{\beta}$.
We model this by the  smoothing function \eqref{eq transition function}
\[
h(y;\bar{\alpha},\bar{\beta}) 
\qquad \mbox{ with,   e.g., $\bar{\alpha}= 0.1, \; \bar{\beta}= 3$ 
}
\]
and set
\[
 \nu_{\mathtt{sf}}(y) = 1 + h\, y^{-\frac{1}{2}}
\]
Solving the equation $ \nu(\frac{a_N}{\mathfrak{a}_0}) a_N = a$  for $a_N$ and solving the relation   $y= \mu(x)\cdot x$   \eqref{constitutive relation mu nu} for $\mu(x)$ leads to
$
\mu(x) = \frac{ 1}{2x}(2x + h^2- h \sqrt{h^2 + 4x} ) $.\footnote{This  $\mu$-function (for $h \equiv 1$) has also been derived in the framework of ``covariant emergent gravity''\citep{Hossenfelder/Mistele:2018}; the authors of the paper confront it with empirical data from galaxy rotation curves (see below).}

The complete interpolation functions of the scalar field model, including shading,   are\footnote{With $\tilde{\alpha}= \bar{\alpha}^2, \; \tilde{\beta}= \bar{\beta}^2$ for the boundaries of the transition interval.}
\begin{subequations}
\beqarr
\nu_{\mathtt{sf}} (y) &=& 1 + h \, y^{-\frac{1}{2}} \, ,  \qquad \hspace{5em} \mbox{with} \; h=  h(x;\tilde{\alpha}, \tilde{\beta}) \label{eq nu-sf} \\
\mu_{\mathtt{sf}} (x) &=& \frac{ 1}{2x}(2x + h^2- h \sqrt{h^2 + 4x} )\, ,  \qquad h=h(x,\bar{\alpha},\bar{{\beta}})\, , \label{eq mu-sf} 
\eeqarr
\end{subequations}
and  thus 
\begin{subequations}
\beqarr
a &=& a_N + h \sqrt{a_N \mathfrak{a}_0} \label{eq a(a-N)-sf} \\
a_N &=& \mu_{\mathtt{sf}}(\frac{a}{\mathfrak{a}_0})\,a = a + \frac{\mathfrak{a}_0}{2}\Big(h^2- h \sqrt{h^2+\frac{4a}{\mathfrak{a}_0}}\,\Big)
\eeqarr
\end{subequations}
The pair $(\mu_{\mathtt{sf}}  , \nu_{\mathtt{sf}}  )$ satisfies the constitutive condition \eqref{constitutive relation mu nu} and the  conditions \eqref{eq asymptotic conditions}.

\section{\small Centrally symmetric case: stars and  simplified  (``round'') models for galaxies \label{section central symmetry}}
Let us consider a centrally symmetric metric with coordinates $x_0= ct,\, x_1=r,\, x_2,\, x_3$,
\beq
ds^2 \underset{Eg}{\doteq} - a(r\,) dx_0^2 + b(r)\,dr^2 + r^2 \big(dx_2^2 + \sin^2 \hspace{-0.2em}x_2\, dx_3^2  \big)\,  \label{eq central symmetric metric}
\eeq
with area radius $r$.\footnote{In our convention   (appendix \ref{subsection appendix central symmetric})  coordinates $x_{\mu}$ are non-metric, thus dimension-less numbers. One might thus prefer to write more precisely $x_0= \{ct\}, x_1=\{r\}$, with $\{ \ldots\}$  signifying the numerical values of a metrical quantity. 
. \label{fn dimensions coordinates}
} 
In this subsection ``$=$'' stands for identity  in the Einstein gauge $\underset{Eg}{\doteq}$.

  The {\em  l.h.s. of the  Einstein equation} \eqref{eq Einstein equ simplified} is in this  metric
\beqarr
G_{00} &=& \frac{a}{r^2}\,\big(1- \frac{1}{b} + \frac{rb'}{b^2} \big)\, , \hspace{12.5em}
G_{11}= \frac{1}{r^2}\, \big(1+\frac{ra'}{a} -b \big) \label{eq G central symmetric} \\
G_{22} &=& \frac{r}{4}\Big(\frac{2\,(a' + ra'')}{a b}  - \frac{r^2 a'^2}{a^2 b} - \frac{(2 ab'+ r a'b')}{a b^2} \Big)\,,  \quad G_{33}= \sin^2 x_2\, G_{22} \, . \nonumber
\eeqarr

\subsection{\small  MOND modification  in the outer region \label{subsection MOND modification of Schwarzschild}}
With regard to the metric  \eqref{eq central symmetric metric} a timelike  vector field is naturally given by  $ (A^{\mu}) = (1,0,0,0) $. In the Milgrom regime the 
 {\em  relativistic Milgrom equation} \eqref{eq core Milgrom equation}   is,  up to a non-vanishing factor,
\[  \sigma'' + \big(\frac{a'}{4a}  - \frac{b'}{2 b} + \frac{1}{r}  \big)\, \sigma' = 0 \, .
\]
It has the solution 
\beq 
\sigma'(r) = a(r)^{-\frac{1}{4}} b(r)^{\frac{1}{2}} \,  \frac{c_1}{r} \,  \label{eq  centrally symmetric scalar field on shell}
\eeq 
with an integration constant $c_1$ (appendix, \ref{subsection appendix central symmetric}). 
For $a(r)\approx 1$ and $b(r)\approx 1$ the solution is close to the  solution $\Phi_{dM}$ of the deep MOND equation in flat space \eqref{eq flat space Milgrom equation}
\[
 \Phi_{dM}'(r)=  \frac{c_1}{r} \, .
 \]
An empirically well confirmed specification  is 
\[
c_1=\sqrt{a_1 M}
\]
 for the central mass $M$, given in length units $M = G c^{-2}\, m$, 
 and $a_1 \approx \frac{H_0}{6}\, c^{-1}$. 
The solution of the full scalar field equation \eqref{eq full Milgrom equation} differs from the r.h.s. of \eqref{eq  centrally symmetric scalar field on shell} by a numerically negligible additive correction only.\footnote{The integral for the full polynomial correction in $|\nabla \sigma|$  can be explicitly evaluated with, e.g., {\sffamily Mathematica}, and allows a precise numerical estimate for the correction.  
}

Using \eqref{eq Theta-bar}  the dominant term of the {\em  r.h.s. of the Einstein equation} becomes
\beqarr
\hat{\Theta}^{(\phi)}_{00} &=& (2- a(r))\, \Box(g)\sigma \, , \qquad \qquad \hat{\Theta}^{(\phi)}_{11} =  \big(\Box(g)\sigma b(r)-2\,\nabla(g)_1 \partial_1 \sigma \big) \label{eq Theta-bar central symmetric} \\
\hat{\Theta}^{(\phi)}_{22}&=& \big( \Box(g)\sigma\, r^2 - \nabla(g)_2\partial_2 \sigma \big) \, ,   \qquad \hat{\Theta}^{(\phi)}_{33} = \sin^2 x_2\, \hat{\Theta}^{(\phi)}_{22} \nonumber
\eeqarr
with
\[
\Box(g)\sigma = \frac{1}{b}\Big(\frac{\sigma''}{b} + \frac{\sigma'}{2}\big(\frac{a'}{a} - \frac{b'}{b} + \frac{4}{r} \big)  \Big) \, ,
\]
which reduces here to the Laplacian with respect to $g$.\\
\noindent
On shell of the  Milgrom equation \eqref{eq  centrally symmetric scalar field on shell} this is 
\beqarr
\Box(g) \sigma  &=&  \frac{c_1}{r^2}\,(a + \frac{r}{4} a')  a^{-\frac{5}{4}} b^{-\frac{1}{2}} \hspace{10em} 
\, , \nonumber \\
\nabla(g)_1 \partial_1 \sigma &=& -3 \frac{c_1}{r^2}\,(a + \frac{r}{4} a')  a^{-\frac{5}{4}} b^{\frac{1}{2}}   \,    \,  \qquad \nonumber   
\eeqarr 
and thus
\beqarr
\hat{\Theta}^{(\phi)}_{00} &=& \frac{c_1}{r^2}\,(a + \frac{5}{4}r a')  a^{-\frac{5}{4}} b^{-\frac{1}{2}}  , \qquad \qquad \hat{\Theta}^{(\phi)}_{11} = 3  \frac{c_1}{r^2}\,(a + \frac{r}{4} a')  a^{-\frac{5}{4}} b^{-\frac{1}{2}}  \label{eq quadratic terms central symmetric}  \\
\hat{\Theta}^{(\phi)}_{22}&=& - c_1  \,(a - \frac{r}{4} a')  a^{-\frac{5}{4}} b^{-\frac{1}{2}}  \, , \quad \;\;\;  \qquad \hat{\Theta}^{(\phi)}_{33} = \sin^2 x_2\, \hat{\Theta}^{(\phi)}_{22} \, . \nonumber
\eeqarr

On shell of the Einstein equation the full energy tensor of the scalar field  \eqref{eq full scalar field energy expression}  has vanishing divergence by the same reasons as in Einstein gravity. This need not necessarily be so for  
 the simplified energy tensor of the scalar field $\hat{\Theta}^{(\phi)}$. In centrally symmetric constellations and on shell of the relativistic  Milgrom equation \eqref{eq core Milgrom equation}  satisfies  the  condition  $\mathtt{div}_1\bar{ \theta} \leq \frac{a_1}{r^2}$; teh divergence  can thus be considered as numerically negligible.\footnote{Details in \citep[sec. 4.1]{Scholz:2025SF} }

\subsubsection*{\small Numerical  solution}
We consider a central symmetric constellation on the level of galaxies with typical mass  $m \sim 10^{11}M_{\odot}$ and distances in $kpc$: 
\beqa
M = m G c^{-2} &=& 10^{-5}\, kpc\,, \;\; a_1= 4.1\cdot 10^{-8}\,kpc^{-1} (\approx  \frac{1}{6} H_1), \;\; \\
 \Lambda &=& 1.4 \cdot10^{-13}\, kpc^{-2} (\approx 3\, \Omega_{\Lambda}\, H_1^2, \; \Omega_{\Lambda}=0.7)\, .
\eeqa

Beyond half the  mean distance of galaxies in clusters the scalar field will be influenced by neighbouring galaxies just as much as by the one at the center. The   range of applicability for the gravitational effects of a galaxy in the region fades out somewhere between $300$ and $600\, kpc$ \citep[tab. 1.1]{Goenner:Kosmologie}. This  fading out will be implemented by a smoothing function similar to \eqref{eq transition function}, 
\[
\tilde{h}(x,\hat{\alpha},\hat{\beta})= 1- g(\frac{x-\hat{\alpha}}{\hat{\beta}-\hat{\alpha}}) \, ,
\] 
where $(\hat{\alpha}, \hat{\beta})=(300,600)$ for the coordinate $r$ will be called the {\em upper transition interval} for galaxies. With the interval
\beq
(\alpha,\beta)=(0.32,1.75) \qquad  \mbox{for the transition function  \eqref{eq transition function}}\;  \label{eq transition parameters model galaxy}
\eeq 
the {\em lower transition region } is characterized by roughly $0.1 \leq \frac{|a_N|}{\mathfrak{a}_0}\leq 3$.

For a 
centrally symmetric  metric \eqref{eq central symmetric metric} ($x_1=r$) the radial acceleration  $a_{rad}$ in coordinate quantities  is given by 
$
\Gamma^{1}_{00} = \frac{a'}{2b} 
$.  Expressed in metrical quantities (adapted to $kpc$ as the metrical unit of length) it is 
\[
a_{rad}= \frac{\sqrt{b}}{a}\,  \Gamma^{1}_{00}\, c^2    \qquad \mbox{(dimension } LL^{-2}L^2T^{-2} = LT^{-2})  \, .
\]
Here it is 
\beq
a_{metr} = \frac{a'}{2a \sqrt{b}} c^2  \, .  \label{eq a-rad metric}
\eeq
The dimensional units of $c$ have to be chosen carefully.\footnote{For coordinate distances $r$ in $kpc$ and  accelerations measured in $cm\,s^{-1}$ one factor of $c$ has to be stated in $kpc\, s^{-1}$, the other one in $cm\, s^{-1}$; i.e., $c^2 \approx  0.2912\, kpc\ cm \, s^{-2} $. }

 For our data  the  Schwarzschild radial acceleration  is close to the MOND acceleration $\mathfrak{a}_0$  at about $r\approx 15.6$ $(kpc)$. We  determine  a numerical solution   in the range $r\geq 10\, (kpc)$ for the first two components of the Einstein equation   \eqref{eq Einstein equ simplified}  with l.h.s. \eqref{eq G central symmetric}   and rh.s. \eqref{eq Theta-bar central symmetric}  (in baryonic vacuum and on shell of the Milgrom equation \eqref{eq core Milgrom equation}) with 
integration constant   \eqref{eq  centrally symmetric scalar field on shell}
\beq
c_1 = \frac{1}{2}\sqrt{a_1M}   \,. \label{eq c-1}
\eeq
 The  factor $\frac{1}{2}$  is necessary to compensate for the factor 2 in \eqref{eq total potential Newton approximation} and \eqref{eq Newtonian mass equivalent phi 2}.

 For initial data for $a(r_0)$ and $b(r_0)$ like in the Schwarzschild solution at $r_0=10\, (kpc)$  the numerical integration of the  coefficients $a(r),\, b(r)$   leads to values not far from the Schwarzschild metric (as to be expected), see figure \ref{fig A B etc model Schwarz}.

\begin{figure}[h]
\includegraphics[scale=0.7]{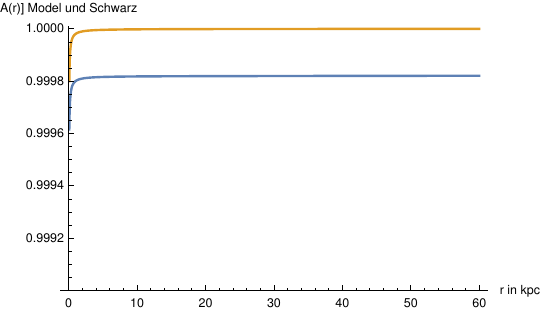} \qquad  \includegraphics[scale=0.7]{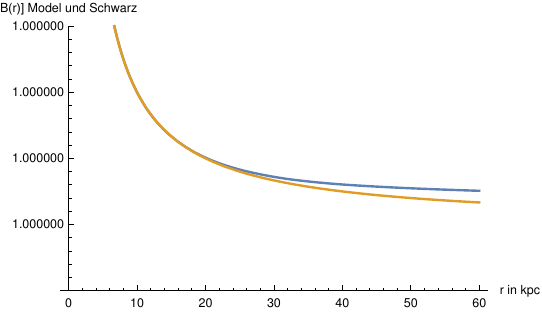} \\[1em]
\caption{\small Comparison of the metric coefficients $a(r)$ and $b(r)$ for the numerical relativistic model (blue) and the Schwarzschild solution (yellow) for a typical galaxy ($m \sim 10^{11}M_{\odot}$). 
\label{fig A B etc model Schwarz}}
\end{figure}

We can now compare 
  the acceleration \eqref{eq a-rad metric} of the  
numerical solution of the relativistic model with  a  usual MOND model with the  simple interpolation function of    \eqref{eq simple interpolation functions}%
 for $n=1$,\footnote{The acceleration $a_M$ of this  MOND model is 
$ a_{M}= \frac{1}{2} a_N + \sqrt{a_0\, a_N + \frac{1}{4} a_N^2} $. 
}  
\[
\mu_M(x)=\frac{x}{1+x} \,, \qquad  \nu_M(y)= \frac{1}{2} \big(1 + \sqrt{1+ \frac{4}{y}}  \big) \, .
\] 
 The    radial acceleration of the relativistic model deviates from  the  classical MOND model for $r< 35\, kpc$ and coincides with it beyond this radius (figure \ref{fig a-rad model MOND}).

\begin{figure}[h]
 \includegraphics[scale=0.8]{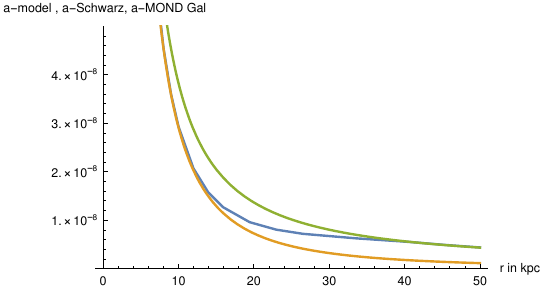} \; \includegraphics[scale=0.8]{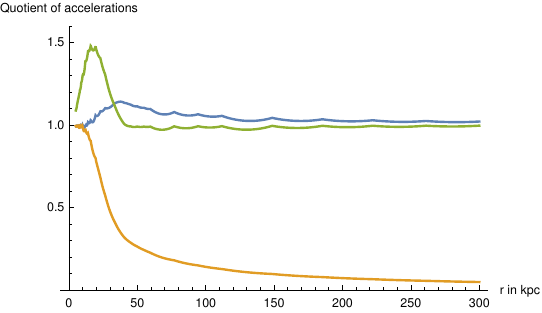} 
\caption{\small Left:  Radial accelerations in $cms^{-2}$ for galaxies ($M=10^{-5}\, kpc$) for the numerical     model  $a_{rel}(r)$   (blue), classical  MOND $a_M$ (green), and Schwarzschild-Newton dynamics  (yellow). Right: Quotient of model accelerations $a_X/a_{rel}$, with $a_{rel}$ the relativistic model, $X$=Newton-Milgrom approximation of relativistic model (blue), $X=$ classical MOND (green), $X=$ Schwarzschild/Newton (yellow). \vspace{1.5em}
 \label{fig a-rad model MOND}}
\end{figure}

For $r\leq 15\, kpc$ the model acceleration is at first identical, later still close to the Schwarzschild acceleration; it supersedes the latter increasingly for larger $r$. 
For $r \geq 30\, kpc$) it is not far from 
 the MOND acceleration  $a_M$ with the ``simple'' interpolation function. 
Figure  \ref{fig a-rad model MOND}, left, shows that  for $r \leq 30\, kpc$ 
the  acceleration of the relativistic model is slightly below $a_M$; later it approaches  $a_M(r)$ increasingly close (both approximate the deep MOND relation).  Figure 5, right,  displays the radial accelerations of the Newton-Milgrom approximation   $a_{M}$   (blue), the simple MOND model (green), and the Schwarzschild acceleration $a_{schw}$ (yellow), effectively identical with the Newton acceleration, all of them divided by $a_{rel}$.  

According to  \eqref{eq Theta-bar central symmetric} the mass-energy density of the scalar field is here
  \beq
\rho_m^{(\phi)} =  (8\pi \varkappa)^{-1} c^{-2} \,(2-a ))\Box(g)\sigma \sim \frac{c_1}{r^2} \,, \label{eq energy density phi central symmetric}
\eeq
 the density of the Newtonian mass equivalent \eqref{eq Newtonian mass equivalent phi 2}  twice as much, $\rho_N^{(\phi)}= 2 \rho_m^{(\phi)}$. Both follow   an inverse square law.

\subsection{\small  Schwarzschild solution in the inner region \label{subsection Schwarzschild}}
Using the flat space Milgrom approximation $a_{\mathsf{rad}}^{(\phi)}(r)= \frac{c^2}{r}\,\sqrt{M\, a_1}$ the boundaries of the transition interval \eqref{eq boundaries transition region} can now expressed in terms of radial distances
\[
 r_0 \leq r \leq r_1 \qquad \mbox{with $r_0, \, r_1$ such that } a_{\mathsf{rad}}^{(\phi)}(r_0) =\beta\, \mathfrak{a}_0\, , \; a_{\mathsf{rad}}^{(\phi)}(r_1) = \alpha\, \mathfrak{a}_0
\]

  For radial distances below $r_0$ 
   the equation \eqref{Weylgeometric EEq} reduces to the classical  Einstein vacuum equation,  its solution to the Schwarzschild metric with central mass $M$. For distances above $r_1$ the  Milgrom regime dynamics prevails. 
   For the  round galaxy model of the last subsection the transition region is  (approximately) given by
\[ 5\, kpc \; \leq \;  r  \, \leq \; 15\, kpc \, .
\] 

\noindent
 At the level of stellar masses, e.g. for the solar system with  $  M \approx  1.97\, AU$ (corresponding to  $m=1\, M_{\odot}$), the transition region is similarly
\[
 2300 \, AU \, \leq \, r \, \leq \, 7000 \, AU  \approx 0.03\, pc\, .
\]

The Milgrom regime for a star of solar mass starts  at about $r \sim 7000\, AU$, while for radial distances below roughly $2.3 \cdot10^3\, AU$ the scalar field is inert (non-dynamical) and the gravitational dynamics is given by the Schwarzschild solution. 
This shows that our {\em  planetary system}  {\em lies deeply inside the Einstein regime}.  
In particular Einstein's calculation of the Mercury perihelion advance remains unchanged.

\section{\small Galaxies \label{section galaxies}}
Galaxies have a more  complicated shape than the spherically symmetric model considered in  
the last section. 
 For  elliptical galaxies the ``round'' model may be used for an overall estimates of the scalar field halo, and  even for a  flat disk  the halo far away from the disk is sufficiently round for taking the central symmetric model as a first approximation for the estimate of its  total energy content.\footnote{For the scalar field halo of a Kuzmin disk see \citep[sec. 4.2]{Scholz:2025SF}}. Before discussing  this let us first shed a glance at empirically well researched constellations:  the radial acceleration relation supported by observational data for a large number of galaxies   and at the Milky Way.

\subsection{\small Radial acceleration relation of galaxies \label{subsection radial acc relation}}
In this subsection the symbol $a$ will be used for the modulus of the total radial acceleration in galaxies,   $a_{bar}$ stands for the (Newtonian) radial acceleration  $|a^{(bar)}|$ due to baryonic matter. 
It is an interesting empirical observation  that the dependence of the total acceleration $a$ on $a_{bar}$  for galactic rotation curves does not essentially depend on the type or the size of the galaxy \citep{McGaugh-et-al:2016}. Over a large range of galactic sizes the dependence is given by a ``universal''  function $a_{bar} \mapsto a(a_{bar})$  (universal in the sense of independence of the galaxy studied). This is a rigid empirical constraint for dark matter halos and also  constrains the interpolation function $\tilde{\nu}$, respectively $\nu$ \eqref{eq mu nu MOND defined}. In the MOND literature it is  called the {\em radial acceleration relation}.

In a detailed empirical study  of 2700 data pairs $(a_{bar},a)$ for 153 galaxies McGaugh and colleagues established the following {\em empirical fit} $a_{emp}$ of this function :
\[
a_{emp}(a_{bar}) = a_{bar}\big(1- e^{-\sqrt{\frac{a_{bar}}{\mathfrak{a}_0}}}  \big)^{-1} \qquad \mbox{resp.} \quad  \tilde{\nu}_{emp}(y)= y\, \big(1-e^{-\sqrt{y}}  \big)^{-1} \quad  \mbox{for} \; y=\frac{a_N}{a_0}\, .
\]
The series expansion in the variable $z=\sqrt{y}$ about $z_0=0$ up to order 3 
\[
\nu_{emp}(y)= \sqrt{y} + \frac{1}{2}y + \frac{1}{12}y^{\frac{3}{2}} + \mathcal{O}(\sqrt{y}^4) \, .
\]
\noindent
After multiplying with $a_0$ this empirical fit boils down to the approximation
\beq
\nu_{emp}(a_{bar}) = \sqrt{\mathfrak{a}_0\,a_{bar}} + \frac{1}{2}a_{bar}  + \ldots \, . \label{eq rar emp order 3}
\eeq
In the Milgrom regime  this differs  from our \eqref{eq a(a-N)-sf}  essentially in the factor $\frac{1}{2}$ in front of $a_{bar}$.\footnote{ Hossenfelder/Mistele's  radial acceleration relation in  \citep{Hossenfelder/Mistele:2018}   agrees exactly with our \eqref{eq mu nu MOND defined} in the Milgrom regime, but does  not foresee a transition to Einstein gravity for stronger gravitation.  \label{Hossenfelder-Verlinde}
} 
The   difference between the two relations may be inspected in figure \ref{fig rar}; it  does not matter deeper inside the Milgrom regime but seems to be  of empirical  relevance  in the transition region and  the  beginning of the Milgrom regime ($ 3 \, \mathfrak{a}_0 \,  > \, a_{bar} > \, 0.1\,\mathfrak{a}_0$), see figure \ref{fig rar}. 
It seems to be important for the the rotational velocities in the Milky Way (see subsections \ref{subsection Milky Way}, \ref{subsection criticism}).

\begin{figure}[h]
\hspace*{7em}\includegraphics[scale=1]{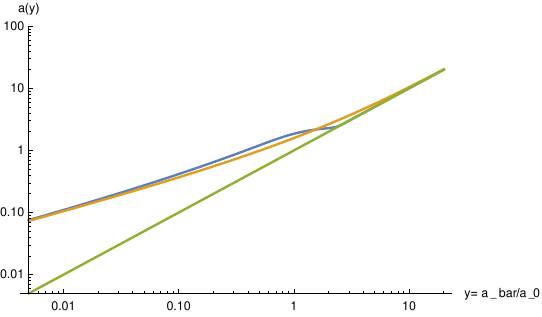}\; 
\caption{\small  Radial acceleration relation $a(y)$ (in multiples of $\mathfrak{a}_0$) with $y=\frac{a_{bar}}{\mathfrak{a}_0}$  for the empirically fitted function   $a_{emp}(y)= \nu_{emp}(y)= y\, \big(1-e^{-\sqrt{y}}  \big)^{-1}$ (orange) from \citep{McGaugh-et-al:2016},  compared with  the respective relation $a_{\mathtt{sf}} (y)$ of the scalar field model for $\tilde{\alpha}=0.1,\, \tilde{\beta}=3$, (blue). The  Newtonian assumption $a(y)=y$  has been  added in green. \label{fig rar} 
}
\end{figure}


\subsection{\small Rotational velocities in the Milky Way \label{subsection Milky Way}}
 Particularly detailed observational data  are provided for our own galaxy  by Ou et al. \citep{Ou-et-al:2023}. It has  been claimed  in \citep{Chan-et-al:2023}  that these observations  ``almost rule out the MOND phenomenology''. This is,  however, not  the case  for the present model (and thus not for MOND phenomenology in general).\footnote{The authors of \citep{Chan-et-al:2023} use a problematic baryonic mass profile of their own and use standard MOND interpolation functions (see subsection \ref{subsection criticism}).
}

The authors of \citep{Ou-et-al:2023} introduce an elaborate model of baryonic matter in the Milky Way, made up from 6 components. For the star mass they combine a bulge model and a disk model,  they  add two dust components (cold and warm) and two gas components (molecular $H_2$ and atomic $H_1$).\footnote{The type of the component models and observation based parameters  are given in \citep[tab. 2]{Ou-et-al:2023}.} 
The resulting Newtonian accelerations induced by each of the components and of the total baryonic mass are given in  \citep[fig. 4]{Ou-et-al:2023} (the total baryonic velocities are shown in the figure below, orange). This allows them to calculate the radial accelerations $a_{bar}(r)$ in the disk plane and  circular velocities $V_{bar}(r)$ expected from the baryonic matter alone. Finally they fit the dark matter  which is needed to fill the gap to 37  data points $v_{obs}(r_j)$, {\scriptsize $ (1\leq j \leq 37)$}  of  observed  circular velocities (black in the figure below) with two dark matter models (Einasto profile and NFW profile).\footnote{The authors find that an Einasto profile gives a considerably better fit than an NFW profile (parameters in \citep[tab. 3]{Ou-et-al:2023}).}

\begin{figure}[h]
\hspace*{5em}\includegraphics[scale=0.9]{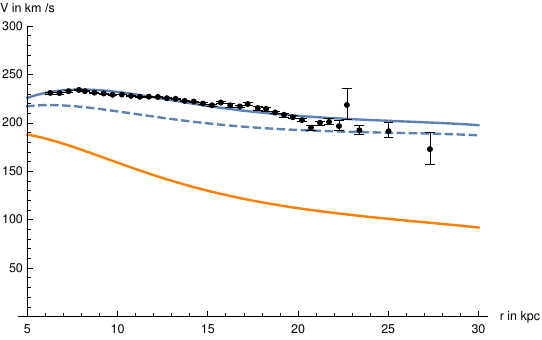} 
\caption{\small Circular velocities in Milky Way.  Black: data points. 
Blue (unbroken): predicted by scalar field approach on the basis of the baryonic density in \citep{Ou-et-al:2023}.
Blue dashed predicted by MOND with interpolation function $\nu_{emp}$ and the same baryonic density.
 Orange: Velocities expected from the baryonic mass only, as modelled in \citep{Ou-et-al:2023}.
\label{fig Milky Way}}
\end{figure}

The radial velocities $V_M$ expected in any MOND approach with interpolation function $\nu$ are easily derived from these data:
\beq
V_M(r)= \sqrt{ \nu\Big(\frac{V_{bar}(r)}{\mathfrak{a}_0 r}\Big) }\, V_{bar}(r) \, 
\eeq

An  evaluation of our scalar field model in the flat space Milgrom approximation  with $\nu(x) = \nu_{\mathsf{sf}}(x)$ like in  \eqref{eq nu-sf} is depicted in figure \ref{fig Milky Way}, blue.\footnote{The baryonic accelerations are derived   with $\bar{\alpha}=0.1, \bar{\beta}=3$ from data  read off from    \citep{Chan-et-al:2023}, fig. 4.}
It shows a  surprisingly  good agreement  with the observational data, in particular compared with other MOND functions. The fit is better than for the interpolation function $\nu_{emp}(y)$ of McGaugh et al. mentioned in the last subsection and much better than for the standard interpolation functions.\footnote{Cf. subsection \ref{subsection criticism}.}

\subsection{\small Scalar field halo in galactic environments \label{subsection halo of galaxies}}
For approximately round galaxies 
 the  energy density  falls off with the inverse square of the central distance \eqref{eq energy density phi central symmetric}. In sufficiently large distances this is also the case for disk galaxies\footnote{For the model of a flat  Kuzmin disk see \citep[sec. 4.2]{Scholz:2025SF} .}
and  results in an approximately linear growth of  the integrated mass-energy content    $M_e^{(\phi)}$ and  the Newtonian  mass equivalent $M_N^{(\phi)}$,  
 \[
 M_e^{(\phi)}(R) \approx  4 \pi   \int_{0}^{R} u^2\, \rho_m^{(\phi)}(u)\, du \, , \qquad 
 M_N^{(\phi)}(R) \approx   4 \pi  \int_{0}^{R}  u^2\, \rho_N^{(\phi)}(u)\, du \, .
 \]
 \begin{figure}[h]
\hspace*{3.5cm}\includegraphics[scale=0.8]{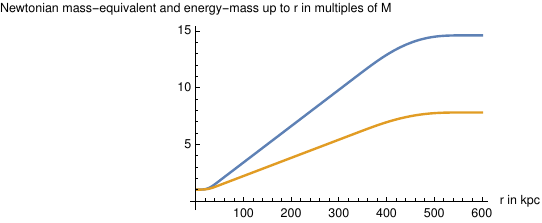}
\caption{\small Total Newtonian mass equivalent of the scalar field $M_N^{(\phi)}$ (blue) and mass expression of the energy density only $M_e^{(\phi)}$ (yellow) up to coordinate distance $r$, expressed in multiples of the central mass $M$.} \label{fig total mass central symm}
\end{figure} \\
\noindent

An estimate for the spherical  case  studied in section 
\ref{subsection MOND modification of Schwarzschild}, here  with typical distances and mass of  galaxies,   can be inspected  in figure \ref{fig total mass central symm}. Its total Newtonian mass equivalent  up to  $500\, kpc$ is about 15  times the central mass $M$ of the galaxy and about 6 times $M$ at $200\, kpc$.  
This is quite considerable and has to be taken into account for the study of cluster dynamics, even in the light of the embedding gravitational field of the cluster gas.

\section{\small Open problems \label{section problems}}
As for any new approach a series of problems remain open for the scalar field model. Here we discuss four of them: gravitational light deflection, galaxy clusters, cosmology, and general criticism of models  with MOND-like dynamics. 

\subsection{\small Light deflection \label{subsection light deflection}}
A particular important application of the reverse Milgrom approximation is the  approximative  calculation  of light deflection in a gravitational field.
The calculation for the standard Newton approximation of pressure-free matter given, e.g., in   \citep{Straumann:GR2004} has to be modified in the Milgrom regime because of the pressure terms of  scalar field.  
  In this subsection we simply set $c=1$.

For a quasi-static metric   in a spacetime $T\times S$ with spatial factor $S$, 
\[
g_{0j}=0\, , \quad \partial_t g_{\mu\nu}= 0\, ,  \qquad    \mbox{and} \quad d\sigma^2=g_{jk}dx^j dx^k  \qquad (j=1,2,3),
\]
the Fermat principle  
\[
\delta \int dt = 0= \delta \int (-g_{00})^{\frac{1}{2}} d\sigma 
\]
implies  that the paths of light in the spatial projection are geodesics of the  ``Fermat metric'' in $S$  \citep[p. 39]{Straumann:GR2004}:\footnote{Light paths with components $\gamma^j(\lambda)$ in $S$ are characterized by the variational principle
$
(\ast)\qquad \delta \int g^{(F)}(\dot{\gamma}, \dot{\gamma})\, d\lambda= 0
$
with fixed end points.}
\[
g^{(F)}= g^{(F)}_{jk} dx^j dx^k\, , \qquad g^{(F)}_{jk} = (- g_{oo})^{-1}g_{jk} \,  
\]

The perturbation  terms $h_{\mu\nu}$ in the weak field metric \eqref{eq weak field metric}
$ g_{\mu \nu} = \eta_{\mu \nu} + h_{\mu \nu}$ 
 may be slightly rewritten by introducing
\[
\gamma_{\mu\nu} = h_{\mu \nu} - \frac{1}{2}h \,\eta_{\mu\nu}, \qquad \quad (h= h_{\nu}^{\nu})
\]
and vice versa
\[
h_{\mu\nu}= \gamma_{\mu\nu} - \frac{ 1}{2} \gamma\, \eta_{\mu\nu} \qquad \quad (\gamma= \gamma_{\nu}^{\nu}) \, .
\]
The weak field Einstein equation, up to 1st order,  is then \citep[p. 220f.]{Straumann:GR2004},
\[
\Box\, \gamma_{\mu\nu} - \eta_{\mu\nu}\partial^{\alpha}\partial^{\beta}\gamma_{\alpha\beta}+ \partial^{\alpha}\partial_{\nu}\gamma_{\mu\alpha} + \partial^{\alpha}\partial_{\mu}\gamma_{\nu\alpha} = 16 \pi G \, T_{\mu\nu}\, .
\]
With respect to the Hilbert gauge with 
$
\partial_{\beta}\gamma^{\alpha\beta}= 0 
$ (which always exist for small perturbations of the Minkowski metric) 
this simplifies to
\beq
\Box\, \gamma_{\mu\nu} = - 16 \pi G\, T_{\mu \nu}\,. \label{diff eq for gamma}
\eeq
and can be solved by retarded integrals ($x\in \R^3$)
\beq
\gamma_{\mu\nu}(x^0,x) = 4 G\, \int_{y\in\R^3} |x-y|^{-1}\, T_{\mu\nu}(x^0-|x-y|, y)\,dy \, .
\eeq
\noindent
In the standard Newton case with a pressure free matter source this implies
\[
\gamma_{00} = 4G\,  \int |x-y|^{-1} T_{00}(t,y)\, dy\,, \qquad \gamma_{jj}= 0 \, .
\]
 and because of $\gamma= - \gamma_{00}$
 \[
 h_{00}= \frac{1}{2}\gamma_{00}\,, \qquad h_{jj}= \frac{1}{2}\gamma_{00}\, .
 \]
A comparison of the Poisson equation for the Newton potential of pressure free matter 
$
\nabla^2 \Phi_N = 4 \pi G\, T_{00}
$
 with \eqref{diff eq for gamma} shows that
$
\gamma_{00} = - 4 \Phi_N \, 
$ 
 thus $ h_{00}=-2\Phi_N, \, h_{jj}= -2\Phi_N$, i.e., 
\beq
g = - (1+2\Phi_N)dt^2 + (1-2\Phi_N)\sum_j dx_j^2  \,, \qquad  g^{(F)}=
\frac{1-2\Phi_N}{1+2\Phi_N} \sum_j dx_j^2  , \label{eq g standard Newton approximation}
\eeq
where  $g^{(F)} \approx (1-4 \Phi_N) \sum dx_j^2$ is the Fermat metric.\footnote{The variational principle $(\ast)$ can be linearized in the length expression  $|\dot{\gamma}|$ as $
\delta \int (1-2\Phi)|\dot{\gamma}|\, d\lambda = 0 $. Then it agrees with the Fermat principle of geometrical optics with refraction index
$
n = 1- 2 \Phi$  \citep[p. 39]{Straumann:GR2004}.
}

 The pressure terms  of the  scalar field energy tensor, even in baryonic vacuum,  make the situation   more involved. From \eqref{eq T-phi-bar} and \eqref{diff eq for gamma}  we get
  \beq
 \Box \gamma_{\mu\nu} = - 2  \big(\Box(g)\sigma (g_{\mu\nu} + 2 A_{\mu}A_{\nu})- 2 \nabla(g)_{(\mu} \partial_{\nu)}\sigma   \big)\, .  \label{eq Box gamma sf model}
 \eeq
  In the quasi-static  case    with flat  metric ($\Box = \nabla^2$) this is at first order:
 \beq
\nabla^2 \gamma_{\mu\nu} = - 2  \big(\nabla^2\sigma (\eta_{\mu\nu} + 2 A_{\mu}A_{\nu})- 2 \nabla_{(\mu} \partial_{\nu)}\sigma  \big) \,   \label{eq gamma quasi static} 
 \eeq
 For the energy component 
$
\nabla^2 \gamma_{00} = - 2 \nabla^2\sigma
 $;   with \eqref{eq Phi-phi} (remember $c=1$ in this subsection)
 \[
\nabla^2 \gamma_{00}= - 2 \nabla^2 \Phi^{(\phi)}\,  \qquad \mbox{and}  \quad   \gamma_{00} = - 2\Phi^{(\phi)}\, .
 \]
 \eqref{eq Box gamma sf model} implies $\Box \gamma = 0$. Ignoring solutions of the wave equation we  conclude 
 \beq
 \gamma=\gamma_{\nu}^{\nu} = 0 \, ,  \qquad \mbox{} \quad h_{\mu \nu}= \gamma_{\mu\nu}\, . \label{eq h and gamma}
 \eeq
  All in all  
  \beq
  h_{00} = -  2\Phi^{(\phi)}\ \qquad \mbox{and}  \quad \nabla^2\,h_{jj} = - 2   \big(\nabla^2 \sigma - 2 \nabla_j\partial_j \sigma  \big) \, . \label{eq hjj scalar field}
 \eeq
 
 This result is considerably more involved than for pressure free (``ordinary'') matter. 
 Only  the time component of the metric perturbation behaves like in the ordinary matter case.  The spatial diagonal perturbations are modified by second order partial derivatives and depend on the spatial constellation of the scalar field 

 In the 
 {\em  centrally symmetric case} (section \ref{section central symmetry}) 
 with  the Minkowski metric in spherical polar coordinates  
 \[
 \widetilde{\eta} = \mathtt{diag}(-1,\,1,\, r^2,\, r^2 \sin^2 \vartheta) \, , \qquad (x^0, x^1,x^2,x^3)= (t, r,\vartheta,\phi)  ,
 \]
 and its perturbation
 \[
 g= \mathtt{diag}(-1+h_{00},\,1+h_{11},\, r^2,\, r^2 \sin^2 \vartheta)\, 
 \]
 \noindent
the Milgrom equation is solved by $\sigma \approx c_1 \ln r$. This follows from   \eqref{eq  centrally symmetric scalar field on shell} which is at first order
  \[
  \sigma' = A^{-\frac{1}{4}}B^{\frac{1}{2}}\, \frac{c_1}{r} \underset{1}{=} (1+\frac{1}{4}h_{00}+ \frac{1}{2}h_{11})\, \frac{c_1}{r} \underset{1}{=} \frac{c_1}{r} \, .
    \] 
  \noindent
 In this case
\[
 \Delta \sigma = \sigma''+\frac{2}{r}\sigma'  =   \frac{c_1}{r^2} =- \sigma'' 
 \qquad \mbox{and} \quad \nabla_1\partial_1\sigma= \sigma'' = - \Delta
 \sigma\, .
 \]

\noindent
From \eqref{eq hjj scalar field}
\[
\Delta h_{11} = - 2 (\Delta \sigma - 2 \nabla_1\partial_1 \sigma)= -6 \,\Delta \sigma = -6\, \Phi^{(\phi)}\,.
\]

 \noindent
  The metric with perturbation from the scalar field only is
 \[
  - (1 +2 \Phi^{(\phi)})dt^2 + (1-6\Phi^{(\phi)})\Big( dr^2 + r^2 \big(d\vartheta^2 + 
  \sin^2\vartheta\, d\phi^2 ) \Big)\, .
 \]
Adding the effects of the baryonic mass, the perturbed metric becomes 
 \beq
 g =  - (1 + 2 \Phi^{(bar)} + 2\Phi^{(\phi)})dt^2 + (1-2\Phi^{(bar)} -6\Phi^{(\phi)})\Big( dr^2 +  r^2 \big(d\vartheta^2 + \sin^2 \vartheta\, d\phi^2 )\Big) \label{eq perturbed metric centr symm}
 \eeq
 and the  Fermat metric 
 \[
g^{(F)} = \frac{1-2\Phi^{(bar)} -6\Phi^{(\phi)}}{1 + 2 \Phi^{(bar)} + 2\Phi^{(\phi)}} \Big( dr^2 +  r^2 \big(d\vartheta^2 + \sin^2 \vartheta\, d\phi^2)\Big) \, .
\]
At first order in the potentials 
\[
g^{(F)} \approx  \big(1- 2(2\Phi^{(bar)}  + 4 \Phi^{(\phi)})\big)\,\Big( dr^2 + r^2 \big(d\vartheta^2 + \sin^2 \vartheta\, d\phi^2) \Big)\, .
 \]
This corresponds  to a refraction index   $1-(2\Phi^{(bar)}+4\Phi^{(\phi)})$.  
 
{\em In the centrally symmetric case      the scalar field potential contributes twice as much} as one would expect from an equally large potential of (pressure free) baryonic matter in  relativistic calculations.\footnote{See also the calculation in \citep[p. 184, box E]{MTW}.}  
  In the MOND literature it is often assumed  that the gravitational light deflection can be calculated from the phantom matter distribution  like for pressure free matter (including the  general relativistic factor 2)  \citep[p. 100]{Famaey/McGaugh:MOND};   but its derivation remains often opaque.   The deviation of our model from the default assumption in MOND ought to have  observable consequences.
  \label{gravitational light deflection}
These  remain to be checked and generalization, most importantly to the axis symmetric case, have to be studied.

\subsection{\small Galaxy Clusters \label{subsection clusters}}
The dynamics of galaxy clusters is a challenge for dark matter theories and an unsolved  problem for MOND \citep{Sanders:2003}. 
For a first exploration of the   scalar field approach to  galaxy clusters we use the  simplifying  assumption of  centrally symmetric shapes for the 
 baryonic mass-density of the hot gas and the star mass, like  in 
  model  calculations in    the astronomical literature  
\citep{Reiprich:Diss}, \citep{Sanders:2003}.  
Given the density functions for the hot gas  $\rho^{(g)}$ and the stellar content $\rho^{(\ast)}$,  
the respective baryonic masses $M^{(g)}$ and $M^{(\ast)}$ up to the central distance $r$,  the    Newtonian  central accelerations $a^{(g,N)}, a^{(\ast,N)}$ and the total Newtonian acceleration due to  baryonic sources  $a^{(bar)}= a^{(g,N)} + a^{(\ast,N)}$  are easily to be  calculated, of course all of them in dependence of $r$.

 The hot gas component of the baryonic mass in  galaxy clusters may be considered s a   smoothly changing external medium $S_g$   for the motion of individual galaxies $S_j$     $(1\leq j \leq N,\; N$ the number of galaxies)  or, alternatively,  for the whole collection of galaxies 
 $S_{\ast} = \bigcup_j S_j$. The reference system for a Newton approximation of the field about each individual galaxy $S_j$ has to be centred on the barycentre of $S_j$. Every galaxy $S_j$ is falling freely in the gravitational field induced by $S_g$ and $\bigcup_{k \neq j} S_k$. 
 We  consider the following alternative assumptions for   the flat space Newton-Milgrom approximations of the field around each galaxy  in such a weakly binding external system:
\begin{itemize}
\item[(A)] The hot gas mass dominates the linear approximation in the cluster barycentric rest system and overlays the  scalar field formation of    
 the galaxies $S_j$. 
\item[(B)] The hot gas induces a scalar field halo  in the cluster barycentric system.  Because of the free fall of the galaxies the scalar field density of each of them \eqref{eq mass-energy scalar field in flat case} may be determined in the  Newton-Milgrom approximation centred on $S_j$ and is exclusively induced by the baryonic matter of $S_j$.  
The effects of the external baryonic masses (gas and the other) are shielded away by the galaxy's free fall.
\end{itemize} 
 Case (A) corresponds to the viewpoint of  the {\em external effect}  (EFE) in classical MOND.  Inside galaxies it  is considered as  empirically well supported for the motion of stars with the corresponding galaxies as their external systems    \citep{Chae-McGaugh-et-al:2020,Kroupa-et-al:2024Star-clusters}. In MOND it is treated as a principle and   assumed to hold also for the motion of galaxies in  galaxy clusters. In the present framework this is not mandatory.\footnote{The different relations between external system and subsystem in MOND and the present scalar field theory may lead to different properties of the linear approximations.}
It even seems to be undermined  by the empirical data for cluster dynamics (see below).

In  case (A) one may work  with the continuous density models  $\rho^{(g)}$ and $ \rho^{(\ast)}$  
of both baryonic mass sources, gaseous and stellar. They  induce scalar field halos satisfying  \eqref{eq flat space Milgrom equation}, respectively  \eqref{eq algebraic transformation a}.
 The  
 accelerations due to the scalar field (here denoted as
  $  a^{(\ldots, \mathtt{sf})}$) are then:
 \begin{itemize}
  \item[(A')] \[
a^{(g,\mathtt{sf})} = h \, \sqrt{\frac{\mathfrak{a_0}}{|a^{(bar)}|}}   a^{(g,N)}  , \; \qquad 
a^{(\ast,\mathtt{sf})} = h \,\sqrt{ \frac{\mathfrak{a}_0}{|a^{(bar)}|} }\, a^{(\ast,N)}\, \qquad \big( h=  h(|a^{(bar)}|,\tilde{\alpha},\tilde{\beta}) \big)
\] 
 \end{itemize}
The total scalar field acceleration $  a^{(\mathtt{sf})}$ is  additively composed,
\[
a^{(\mathtt{sf})} = h\,\sqrt{ \frac{\mathfrak{a}_0}{|a^{(bar)}|} }\, a^{(bar)}  =  a^{(g,\mathtt{sf})} + a^{(\ast,\mathtt{sf})} \, .
\]
Under the assumption of central symmetry  the accelerations can be straightforwardly  converted  into respective mass expressions 
$
 M^{(\mathtt{sf})},   M^{(g,\mathtt{sf})}, M ^{(\ast,\mathtt{sf})} 
$
 ($M= r^2\,G_N^{-1}\, a$) and vice versa. 
 
    The bulk of baryonic matter given by the gas mass is often approximated by a $\beta$ model.\footnote{The density function of a $\beta$ model is  $ \rho(r)= \rho_{0} \Big(1+ \frac{r^2}{r_c^2}  \Big)^{-3 \beta + 1/2} $ with the parameters central density $\rho_0$, core radius $r_c$ and  $\beta$. 
}  The Newtonian mass $M^{(g,N)}$ and acceleration $a^{(g,N)}$ induce a scalar field halo according to \eqref{eq flat space Milgrom equation}, respectively  \eqref{eq algebraic transformation a} with acceleration $a^{(g,\mathtt{sf})}$ and Newtonian mass equivalent  $  M^{(g,\mathtt{sf})}$. The distribution of the star mass is  less regular. Its empirical determination is difficult; reliable values are given in the literature at specific distances only, e.g. for $r_{500}$.\footnote{$r_{500}$ is the distance from the center at which the total dynamical mass density is down to $500\,\rho_{crit}$ (critical cosmic  density).}

 In the case (B) the scalar field halo of the gas can be computed like above (A'). On the other hand, the acceleration induced by the scalar field of the galaxies in a continuity model for the star mas $ \rho^{(\ast)}$ lies   in the Milgrom regime and is given by
 \begin{itemize}
 \item[(B')]
 \[
 a^{(\ast,\mathtt{sf})} = \sqrt{ \frac{\mathfrak{a}_0}{|a^{(\ast,N)}|} \big)}\, a^{(\ast,N)} \, .
 \]
  \end{itemize}

  It is not easy to decide between (A) and (B) on the theoretical level; we therefore check the alternatives against the empirical data for a typical cluster, say Coma.

  Let $M_{500}^{(tot,N)}$  denotes the total dynamical mass of the cluster at the radius $r_{500}$, determined in Newton gravity,    $M^{(gas)}_{500}$ 
 the mass of the hot gas,  $M^{(\ast)}_{500}$  the estimated luminous mass of stars/galaxies and  $M_{500}^{(bar)}= M^{(gas)}_{500}+M^{(\ast)}_{500}$. In  \citep[tab. 1]{Reiprich/Zhang_ea:Corr} we find for Coma  in units of  $10^{13}\, M_{\odot}$   
\beq
M_{500}^{(tot)} = 65.5 \pm 7.9\, , \quad M^{(gas)}_{500} = 8.42 \pm 0.63\, , \quad  M^{(\ast)}_{500} =1.31 \pm 0.18 \, , \label{eq parameters Coma}
 \eeq
with  $r_{500} \approx 1.278\, Mpc$  according to \citep[tab. 4.2]{Reiprich:Diss}.\footnote{Assuming $h_0=\frac{73}{50}$.}
 
The   missing  mass in Newtonian gravity
 \[
  \Delta M^{(N)}_{500} = M_{500}^{(tot)} -   M_{500}^{(bar)} \approx 56 \cdot10^{13}\,M_{\odot} \,  
  \]
is often understood as the amount of  dark matter up to this radius.\footnote{The ratio of the missing mass to the hot gas mass  $\frac{ \Delta M^{(N)}}{ M^{(bar)}} \approx 5.8$ is sometimes called  the {\em mass discrepancy} ratio  at $r_{500}$ in the Newtonian approach \citep{Sanders:2003}.}
 
 The scalar  field contributions to the accelerations at $r_{500}$    are far below $\mathfrak{a_0}$, in units of  $ 10^{-8}\, cm\, s^{-2}$
 \begin{center}
 
 \begin{tabular}{|l|r|r|r|r|r|r|}
\hline 
Case & $ a^{(g,\mathtt{sf})}$ & $a^{(\ast,\mathtt{sf})}$ & $a^{(\mathtt{sf})}$ \\ 
\hline  
(A) & 0.315 & 0.043 & 0.358 \\ 
\hline 
(B) & 0.315 & 0.116 & 0.431  \\ 
\hline
\end{tabular} \\
 \end{center}
 \noindent
The associated Newtonian mass equivalents of the scalar field $ M^{(g,\mathtt{sf})}, \,M^{(\ast,\mathtt{sf})} $,  $M^{(\mathtt{sf})}= M^{(g,\mathtt{sf})},  + M^{(\ast,\mathtt{sf})} $ and the missing masses of the present approach, 
 \[
  \Delta M^{(\mathtt{sf})}_{500} = M_{500}^{(tot)} - ( M_{500}^{(bar)} + M_{500}^{(\mathtt{sf})} ) \, , 
  \]
  can be derived from this. Their values 
 are given in the following table  in  $10^{13} M_{\odot}$.\footnote{In  a discrete representation of the star mass,  $S_{\ast} = \bigcup_j S_j$, the Newtonian mass equivalent of the scalar field about each galaxy  may be estimated to lie  between 6 and 15 times the baryonic mass of the galaxy, depending on the distance to the next galaxy (see \ref{subsection halo of galaxies}). One  would expect the mass equivalent of the star mass  to be roughly 
$ M ^{(\ast,\mathtt{sf})}  \sim \; 10\, M ^{(\ast)}$. This is sufficiently close to the value of $M ^{(\ast,\mathtt{sf})} $ in case (B) for lending credibility to the continuity model of the star mass.  } 

\noindent
\begin{center}
\begin{tabular}{|l|r|r|r|r|r|r|r|r|r|}
\hline 
Case &  $ M^{(g,\mathtt{sf})}$ & $M^{(\ast,\mathtt{sf})}$ & $M^{(\mathtt{sf})}$ &$M^{(\mathtt{bar})}$ & $M^{(\mathtt{bar})} + M^{(\mathtt{sf})}$ &   $\Delta M^{(\mathtt{sf})}$ \\ 
\hline  
(A)  & 37.0 & 5.0  & 42.0 & 9.7 & 51.7 & 13.8\\ 
\hline 
(B)  & 37.0 & 13.6  & 50.6 & 9.7&   60.3 & 5.2\\ 
\hline
\end{tabular} 
\\[0.5em]
\end{center}

This is an interesting result. Already the missing mass of case (A) is  less than half the missing mass of classical MOND calculations.  Assuming MOND dynamics  R. Sanders  arrived  for Coma at an unexplained residuum $\Delta M^{\mathtt{M}}_{500} \approx 30\cdot 10^{13} M_{\odot}$  \citep{Sanders:2003}.  It led him to the hypothesis of an additional component of dark matter (probably made up of   sterile neutrinos). 

Case (B) looks even better. The { missing mass} $\Delta M^{(\mathtt{sf})} = 5.2\cdot 10^{13} M_{\odot} $ for the central values of \eqref{eq parameters Coma} lies  inside the   empirical error interval. {\em This speaks strongly in favour of the hypothesis (B)}. The result has  to be checked seriously  for more galaxies, but that  cannot be done here.\footnote{Data for 19 galaxies (including Coma) are available in \citep{Reiprich:Diss,Reiprich/Zhang_ea:Corr}.}

\subsection{\small Cosmology \label{subsection cosmology}}
Large scale cosmological models of the present approach lie in the Einstein regime of the scalar field.  This may be considered a weak spot or an advantage, depending on the scientific perspective taken. In   the light of the growing difficulties of the actual standard model of cosmology it appears at least dissatisfying that our Lagrangian \eqref{eq Lagrangian phi-2}  does not endow the scalar field  with a dynamical  role at the cosmological level.  Modifications and improvements may be looked for. 

\subsection{\small Criticism of MOND in general \label{subsection criticism}}
Many criticism of MOND in general have  been brought forward in the literature. Some of it (lacking connection to general relativity, indecision whether it modifies dynamics or gravity etc.) become pointless in the present approach. Others  do still apply, most importantly among them,  the non-fundamentality of the approach. This deficiency may perhaps be relaxed  in  case  the Weyl geometric gravity approach in \citep{Ghilencea:2019JHEP,Ghilencea:2022SMWeyl,Ghilencea/Harko:2021,Ghilencea:2023,Ghilencea:2025,Burikham/Harko-ea:2023} progresses and turns out  helpful for a program of integrating   gravity with quantization,\footnote{Steps in this direction among others in 
 \citep{Ghilencea:2025,Percacci:RG_flow,Codello_ea:2013}. } and if the link of the model to the latter can be consolidated.

Moreover, in the   paper  \citep{Chan-et-al:2023} the authors claim empirical inadequacy  of MOND at the level of galaxy dynamics, the very range on which the latter  boasts its main success.  They analyse the recent data of rotation velocities measured in the Milky Way \citep{Ou-et-al:2023} from a classical MOND perspective, with regard to  two families of interpolation functions, the algebraic one  of \eqref{eq simple interpolation functions} and an analytic one  considered in \citep[eq. (53)]{Famaey/McGaugh:MOND}, 
\[
\nu_{\delta}(y)= \Big(1- e^{-\frac{\delta}{2}}  \Big)^{-\frac{1}{\delta}}\, .
\]
 The authors argue that with both types of  interpolation  functions the effective total accelerations calculated in MOND are incompatible with the observational data of   \citep{Ou-et-al:2023}. They draw the conclusion that this ``can almost rule out MOND phenomenology'' in general. 

In subsection   \ref{subsection Milky Way} we have shown that this is {\em not the case} for the present approach. As the interpolation functions of the flat space Milgrom approximation satisfy the defining conditions  \eqref{constitutive relation mu nu} of a MOND algorithm the generalization of the authors is premature. 
Moreover, even for classical MOND the argument of the authors of \citep{Chan-et-al:2023} is  not impeccable. 
They  derive their inconsistency result on the basis of a comparably  simple  baryonic mass model for the Milky Way with two components only, composing a bulge model with a flat disk, rather than   the elaborate model used by the authors of \citep{Ou-et-al:2023}. This is  problematic, as the effective total acceleration of any MOND model is highly sensitive to variation of the Newtonian acceleration of the baryonic mass. 

But even using the baryonic mass distribution of  \citep{Chan-et-al:2023} the standard interpolation functions do not lead to an acceptable reproduction of the observational data.  
\begin{figure}[h]
\hspace*{5em}\includegraphics[scale=0.9]{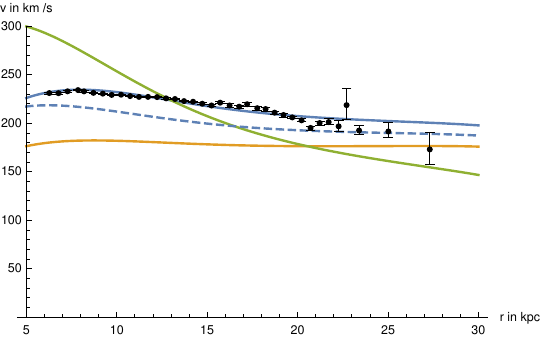} 
\caption{\small Circular velocities in Milky Way.  Black: data points. 
Blue (unbroken): predicted by scalar field approach on the basis of the baryonic density in \citep{Ou-et-al:2023}.
Blue dashed: predicted by MOND with interpolation function $\nu_{emp}$ and the same baryonic density.
 Orange: predicted by MOND with standard interpolation function $\nu_n, \, n=1$. \\ Green: predicted by MOND with standard interpolation function $\nu_{\delta}, \, \delta=1$.
\label{fig Milky Way Chan}}
\end{figure}

 In this sense, the remarks of Chan et al. hint to an important observational domain (rotation curves of the Milky Way with baryonic accelerations close to the critical value $\mathfrak{a_0}$) for which the present model outmatches the two families of MOND interpolation functions mentioned above. This has to be checked further.

\section{\small Resum\'ee and discussion \label{section resumee}}
 Already at the theoretical level we have found   interesting properties of the   scalar field  introduced in this paper (section \ref{subsection Lagrangian}). First and above all, it has been shown that a single gravitationally coupled scalar field is able to  generate MOND-typical free fall trajectories in the flat  approximation and is accompanied by  gravitational light  bending  resembling
  the one expected in a dark matter approach or in ordinary MOND, although not identical with it. 
This is no triviality; earlier approaches to relativistic generalizations of MOND have postulated quite involved  structures, often with two basically unrelated Riemannian metrics and diverse  additional fields (section \ref{subsection other relativistic MOND}). In the present approach the metrical generalization of Riemannian geometry is achieved by  the weakest possible variant of Weyl geometry: the integrable case, with the Einstein gauge (frame) specifying physical measurements. On the one hand, the scalar field is part and parcel of the Weyl geometric modification of  Riemannian geometry, i.e., of the gravitational structure. On the other side its energy-momentum tensor entails an  important addition  to the right hand side of the Einstein equation, letting it appear as a peculiar type of dark matter (with unexpected pressures).  In this sense it fits seamlessly into the persepctive of \citep{Lehmkuhl/Martens:2020}. 
For strong gradients the scalar field is screened away (Einstein-Newton regime);  only for weak gradients the scalar field modifies Einstein-Newton gravity (Milgrom regime). 

For the central symmetric  case these effects can be quantitatively investigated by  numerical calculations showing how the  Schwarzschild  metric is deformed in the Milgrom regime, and  how the  energy and Newtonian mass equivalent  of the scalar field add up in the long range (section \ref{subsection MOND modification of Schwarzschild}).  

 In the flat space approximation the general relativistic dynamical equation of the scalar field specializes to the non-linear Poisson equation known from  MOND (deep MOND case),  and the Newton approximation of the Einstein equation acquires a non-negligible source term due to the scalar field in addition to the baryonic sources (section \ref{section flat approximation}). 
 The flat approximation can be used to infer backwards crucial properties of the scalar field, in particular those relating to the energy-momentum tensor. In the Milgrom regime the present model has  precisely determined ``interpolation'' functions in the sense of the MOND approach,  which facilitate a  quantitative comparison with other MOND models (section \ref{subsection MOND interpolation functions}). 

At the galactic level the  radial acceleration function implied by the interpolation functions of the present model is close to the one  empirically determined  by McGaugh et al. \citep{McGaugh-et-al:2016}. Relevant differences may arise  close to the critical  acceleration for MOND $\mathfrak{a_0}$ (section \ref{subsection radial acc relation}). Observational data on radial velocities in the Milky Way give precise data on this acceleration regime
 \citep{Ou-et-al:2023}. Starting from  the data  given there on the baryonic matter distribution in the Milky Way  the scalar field model predicts the total accelerations and the radial velocities surprisingly well (section \ref{subsection Milky Way},  \ref{subsection criticism}). Moreover, for galaxy clusters it may even make a crucial difference to the classical MOND calculations: If one adds the Newtonian mass equivalent of the scalar field halos of the freely falling galaxies to that of the hot gas mass, the total dynamic mass of the model seems to suffice for explaining the observational data; at least this is so for Coma (section \ref{subsection clusters}). 

 A major open problem results from the  pressure term of scalar field energy tensor. It has strong repercussions on the gravitational light deflection expected in the model. The contribution of the scalar field to the 
 refraction index 
 differs from the one expected for classical MOND. In the centrally symmetric case it  is twice the one expected there and for cold dark matter (section \ref{subsection light deflection}). This should allow to discriminate between the approaches. If the  present model fails in this point it will be empirically refuted; the theoretical contributions mentioned above would then remain the only achievements of the  approach.
\\[0.8em]

\noindent
\textbf{Acknowledgements}: I thank P. Kroupa  for comments on an earlier version of this paper. An anonymous referee made me aware of the importance of  \citep{Ou-et-al:2023,Chan-et-al:2023}.\\


\section{\small Appendix \label{section appendix}}
\subsection{\small Short outline of Weyl geometric methods and the notation used  \label{appendix Weyl geometry}}
A {\em  Weylian metric}  may be characterized by an equivalence class of pairs $[(g,\varphi)]$ consisting of a semi-Riemannian metric $g=g_{\mu\nu} dx^{\mu}dx^{\nu}$, the {\em Riemannian component} of the Weylian metric,  and an  associated  1-form  $\varphi= \varphi_{\nu}dx^{\nu}$ representing a {\em scale connection}. Equivalences arise from  point dependent rescalings of the Riemannian component by a positive function $\Omega(x)$,
\[
\tilde{g}= \Omega^2  g \, ,
\]
 and an associated gauge transformation of the 1-form, 
\[
 \tilde{\varphi}=  \varphi- d \log \Omega  \qquad (\log = \ln)  \, .
\]
Choosing a representative $(g,\varphi)$ of the class is the same as gauging the Weylian metric.  
The latter has  a unique compatible {\em scale invariant affine connection} $\Gamma$ with an  associated covariant derivative $\nabla = \nabla(\Gamma)$. 
One may write the Weylian affine connection and its covariant derivative of a vector field $X$ in the form
\beq
\Gamma = \Gamma(g) + \Gamma(\varphi) \qquad \mbox{respectively} \qquad \nabla {\nu} X^{\mu } = \partial_{\nu}X^{\mu}+ \Gamma_{\nu \lambda}^{\mu}X^{\lambda} \label{eq decomposition affine connection}
\eeq
where $\Gamma(g)$ denotes the Levi-Civita connection of $g$ and $\nabla(g)$ the associated covariant derivative.  $\Gamma(\varphi)$ is an expression (not a connection but a tensor) codifying the contribution of $\varphi$ to the scale invariant affine connection.  Sometimes  $\Gamma$ is  written as $\Gamma(g,\varphi)$ for emphasizing the combination of the gauge dependent contributions, although this hides its gauge independence.  

 For a  scale covariant field $X$  of (Weyl-)  weight $w$  transforming under rescaling  by
$
\tilde{X}= \Omega^w X 
$  $\nabla X$ is not scale covariant, while 
\beq 
D_{\nu} X = \nabla(g)_{\nu}X + w \, \varphi_{\nu}\, X \, 
\eeq
is again scale covariant of  the same weight $w$. $D$ is called the {\em scale covariant derivative} operator. 
It satisfies the condition
\beq
D_{\nu}g_{\mu\lambda} = 0 \, ;\label{eq Dg=0} 
\eeq
in other words,  it is  compatible with the metric  in the sense of Weyl geometry.  

The curvature tensors $Riem = Riem(\Gamma)$ (generalized Riemann curvature) and  $Ric(\Gamma)$ (Ricci) derived from the affine connection $\Gamma$ are  scale invariant. In any gauge they can be composed similar to $\Gamma$, e.g.,  $Riem(g) + Riem(\varphi)$ etc. The scalar curvature $R(g,\varphi)$  depends on the gauge and is scale covariant of weight $-2$.
For more details see  appendix \ref{appendix useful formulas} and the classics  \citep{Weyl:GuE,Weyl:InfGeo,Pauli:1921,Eddington:Relativity,Dirac:1973}.\footnote{More recent presentations of Weyl geometry can be found in 
\citep{Blagojevic:Gravitation,Drechsler/Tann,Perlick:Diss} (difficult to access), for selected aspects  \citep{Codello_ea:2013,Ohanian:2016}. Integrable Weyl geometry is presented in \citep{Dahia_ea:2008,Romero_ea:Weyl_frames,Almeida/Pucheu:2014,Quiros:2014a,Scholz:2011Annalen}.  Expressions for Weyl geometric derivatives  and  curvature quantities are derived in  \citep{Yuan/Huang:2013,Miritzis:2004,Ghilencea:2025}, more  mathematical  perspectives in  \citep{Folland:1970,%
Calderbank/Pedersen:1998,Gauduchon:1995,Higa:1993,Ornea:2001,Gilkey_ea}.}

Here we work mostly with Weylian metrics of dimension $n=4$ and Lorentzian signature $(-+++)$. 
Moreover, in our context the curvature of the scale connection is  assumed to vanish,
   \[
d\varphi = 0 \, ,
\]
while the structure is enriched by a real valued {\em scale covariant field} $\phi$ of weight $-1$.

There are two special gauges. One exploits the vanishing of the scale connection,  $\varphi=0$,  and characterizes the Weylian metric  by its  Riemannian component only; this is called the {\em Riemann gauge}.  Another distinguished gauge is the one in which the scalar field is constant, $\phi(x)=\phi_0$ (for all $x$) and may be called {\em  scalar field gauge}. As the present model assumes  non-minimal coupling of $\phi$ to  gravitation,  this gauge has strong analogies to the Einstein frame in JBD theory;  accordingly it will  also be referred to as the {\em  Einstein gauge} of the Weylian metric. {\em Matter fields are  assumed to break the scale symmetry and to couple with gravity in the Einstein gauge.} Thus test matter   follows the geodesics of the $\Gamma(g_E)$ with $g_E$ the Riemannian component of the  Einstein gauged metric.

Equalities  in these gauges  are  denoted 
$
\underset{Rg}{\doteq} \; \mbox{for the Riemann gauge}, \;  \underset{Eg}{\doteq} \; \mbox{for the Einstein gauge}.
$
\noindent
In Riemann gauge the scalar field  is written in exponential form,
\beq
\phi(x) \underset{Rg}{\doteq} \phi_0 e^{-\sigma(x)}\, . \label{eq phi exponential form}
\eeq
Because the transition from Riemann to Einstein gauge is accomplished by  rescaling with  $\Omega=e^{-\sigma}$, the scalar field and the 
 scale connection in Einstein gauge are  given by 
\[
\phi \underset{Eg}{\doteq} \phi_0 \qquad \mbox{and} \qquad 
\varphi \underset{Eg}{\doteq}  - d \log \Omega = d \sigma \,. 
\]
This shows  that the scalar field $\phi$ and the integrable scale connection $\varphi$ can be traded against each other. 
 Taken together  they represent  the scalar field degree of freedom of the model.  

 No direct physical relevance is assumed  in the following for the scale invariant affine connection $\Gamma$; the Weyl geometric framework is mainly used as a    symbolic tool particularly well adapted for forming scale invariant Lagrangians and the derivation of scale covariant dynamical equations. Reference to empirical data  will be made via the Einstein gauge. In this sense the present model is placed in a kind of overlap region  of JBD theory and Weyl geometric field theory. Even though it uses the latter in  a moderate way, it cannot be reduced to  JBD theory. The latter confines itself to shifting  perspectives in a  family of conformally related Riemannian metrics endowed with a rescaling field, but lacks a consistent framework for  working with scale covariant derivatives and dynamical equations. 

Similar to \eqref{eq decomposition affine connection} for the affine connection the  curvature quantities of Weyl geometry in any given gauge $(g,\varphi)$  can be composed from a Riemannian component (the respective quantity with regard to $g$ only) and an additional  term due to  the  scale connection:\footnote{\citep[GA II, p. 21]{Weyl:InfGeo}, but note that Weyl's scale connection is twice ours. For more recent literature see, e.g.,  \citep[sec. 2.1]{Ghilencea:2025}.} 

\beqa
Riem(g,\varphi) &=& Riem(g)+ Riem(\varphi) \hspace{3.5em} \mbox{(curvature tensor)} \\
G_W = G(g,\varphi) &=& Riem(g,\varphi) - \frac{1}{2}R(g,\varphi )\, g \qquad \mbox{(Einstein tensor)}\\
  &=& G(g) + G(\varphi) \\
  R(g,\varphi) &=&  R(g)+R(\varphi)  \hspace{3.5em} \mbox{(scalar curvature)} 
\eeqa
The first two are scale invariant, while $R(g,\varphi)$ is scale  covariant of weight $-2$.

{\em  Scale connection contributions to Weyl geometric terms} in dimension $n$ are: 
\begin{subequations}
\beqarr 
\Gamma(\varphi)_{\mu \nu}^{\lambda} &=& \delta_{\mu}^{\lambda}\varphi_{\nu} +   \delta_{\nu}^{\lambda}\varphi_{\mu}  - g_{\mu\nu}\,\varphi^{\lambda}  \\
Ric(\varphi)_{\mu \nu} &=& (n-2)\big(\varphi_{\mu}\varphi_{\nu} - \nabla(g)_{(\mu}\varphi_{\nu})\big) -\big((n-2)\varphi_{\lambda}\varphi^{\lambda} + \nabla(g)_{\lambda}\varphi^{\lambda}  \big)\, g_{\mu \nu}    \\
R(\varphi) &=& - (n-1)(n-2)\varphi_{\lambda}\varphi^{\lambda} - 2(n-1)\nabla(g)_{\lambda}\varphi^{\lambda}   \label{eq Weyl geom curvatures} \\
G(\varphi)_{\mu \nu} &=&  (n-2)\Bigl( \varphi_{\mu} \varphi_{\nu}-  \nabla(g)_{(\mu}\varphi_{\nu)} + \big(\frac{1}{2}(n-3)\varphi_{\lambda}\varphi^{\lambda} + \nabla(g)_{\lambda} \varphi^{\lambda} \big)g_{\mu\nu}  \Bigr) \label{eq Weyl geom Einstein tensor} \\
\mbox{for $n=4$}: &= & 2 \bigl( \varphi_{\mu} \varphi_{\nu}-  \nabla(g)_{(\mu}\varphi_{\nu)} \bigr) + \big(\varphi_{\lambda}\varphi^{\lambda} + 2\nabla(g)_{\lambda} \varphi^{\lambda} \big)g_{\mu\nu}  \nonumber \\
 &\underset{Eg}{\doteq}&  2 \bigl( d\sigma_{\mu} d\sigma_{\nu}-  \nabla(g)_{(\mu}d\sigma_{\nu)} \bigr) + \big(d\sigma_{\lambda}d\sigma^{\lambda} + 2\nabla(g)_{\lambda} d\sigma^{\lambda} \big)g_{\mu\nu} \label{eq sigma contribution Einstein tensor}
\eeqarr
\end{subequations}

\subsection{\small Variational derivatives \label{appendix variations}}
Variational derivatives of a Lagrangian density 
\[
\mathcal{L} = L \sqrt{|g|}
\]
 with regard to a field $Y$ 
are  written  as Euler expressions denoted by square brackets with lower index $Y$,
\[ 
{[}\mathcal{L}]_Y  = 
\frac{\delta \mathcal{L}}{\delta Y}= \frac{\partial \mathcal{L}_Y}{\partial Y} - \partial_{\nu} \frac{\partial \mathcal{L}_Y}{\partial(\partial_{\nu } Y)} \pm \ldots \, ,
 \, 
\]
with  the Euler equation ${[}\mathcal{L}]_Y =0$, the dynamical equation for $Y$. \\[-0.3em]

\noindent 
\textbf{Variation  $\delta g^{\mu \nu}$}\\[0.2em]
The variation  $[\mathcal{L}_H]$ of the Weyl geometric scalar curvature with a non-minimally coupled scalar field leads to \citep[eq. (2.17)]{Drechsler/Tann}\footnote{A  detailed derivation in  \citep[eq. (220)]{Tann:Diss}.} 
\beq {[}\mathcal{L}_H  ]_{g^ {\mu\nu}} = \Big( \frac{(\hbar c)^{-1}}{2} (\xi \phi)^2  G(g,\varphi)_{\mu\nu} + \frac{(\hbar c)^{-1}}{2}\big((D_{\lambda}D^{\lambda}(\xi \phi)^2)\,g_{\mu\nu} - D_{(\mu}D_{\nu)}(\xi\phi)^2  \big) \Big)\,\sqrt{|g|} \, , \label{eq variation Hilbert term}
\eeq 
with   the  Weyl geometric Einstein tensor  $ G(g,\varphi) = Ric(g,\varphi)-\frac{1}{2}R(g,\varphi) g$. 
  The variational derivative   \eqref{eq variation Hilbert term}
contains two contributions not known from the Riemann-Einstein case. 
One  results from $G(\varphi)$,  the contribution  of the scale connection to the Weyl geometric curvature  \eqref{eq Weyl geom Einstein tensor}, the other one from 
the non-minimal coupling of the scalar field like in  JBD theory \citep[eqs. (3.1), (3.5)]{Capozziello/Faraoni}, \citep[(2.1), (2.6))]{Fujii/Maeda}. Here it appears in a scale covariant form (the second term on the r.h.s in \eqref{eq variation Hilbert term}). 
Taken together the second order terms cancel in  dimension $n=4$. In the Einstein gauge they are with  \eqref{eq Weyl geom Einstein tensor},\eqref{eq Theta-phi-H in Einstein gauge}: 
\beq
(\xi \phi)^2 G(\varphi)_{\mu \nu} +   \big((g_{\mu \nu}\, D_{\lambda}D^{\lambda}(\xi \phi)^2 - D_{(\mu}D_{\nu)}(\xi\phi)^2  \big)   \underset{Eg}{\doteq} (\xi \phi_0)^2 \, \big(3\partial_{\lambda}\sigma \partial^{\lambda}\sigma\, g_{\mu \nu} - 6 \partial_{\mu}\sigma \partial_{\nu}\sigma \big)  \label{eq G(varphi) and additional term}
\eeq
\noindent

 The variation of all scale invariant Lagrange densities leads to scale invariant 
 expressions with  Weyl weight $-2$  and of   physical dimension $EL^{-1}$  (metrical only in  the Einstein gauge).

\beqarr  
(\hbar c)\sqrt{|g|}^{\,-1}  [\mathcal{L}_H]_{g^{\mu \nu}} &=& \frac{(\xi \phi)^2}{2}  G(g,\varphi)_{\mu\nu}\, + \frac{1}{2}\big(g_{\mu\nu}\, D_{\lambda}D^{\lambda}(\xi \phi)^2 - D_{(\mu}D_{\nu)}(\xi\phi)^2  \big)\, \nonumber \\ 
(\hbar c)\sqrt{|g|}^{\,-1}  {[} \mathcal{L}_{\phi_2} {]}_{g^{\mu \nu}} &=& - \frac{\alpha}{2} \xi^2 D_{\mu}\phi D_{\nu}\phi  - \frac{1}{2} {L}_{\phi_2} g_{\mu \nu} \label{eq scale covariant Euler exp g} \\
(\hbar c)\sqrt{|g|}^{\,-1}  {[} \mathcal{L}_{\phi_3} {]}_{g^{\mu \nu}} &=& - \, s_{\phi}\,\, \frac{\beta}{2} \xi^3 \phi^{-2}|D \phi| \,D_{\mu}\phi \,D_{\nu}\phi  - \frac{1}{2}  {L}_{\phi_3}\, g_{\mu \nu}  \nonumber \\
(\hbar c) \sqrt{|g|}^{\,-1}  {[} \mathcal{L}_{_2\phi} {]}_{g^{\mu\nu}} &=&  \xi^2\phi\, \Big( \big( D_{\mu}D_{\nu}\phi  - \frac{1}{2}D_{\lambda}D^{\lambda}\phi\,g_{\mu\nu}  \big)\,A_{\lambda}A^{\lambda}  +   D_{\lambda}D^{\lambda}\phi \, A_{\mu}A_{\nu}   \Big)\,    \nonumber \\
\sqrt{|g|}^{\,-1}    {[} \mathcal{L}_{V} ]_{g^{\mu \nu}} &=&  - \frac{1}{2}  {L}_{V} g_{\mu \nu}  \nonumber 
\eeqarr
For baryonic matter we  consider  the Einstein gauge
\[
 \sqrt{|g|}^{\,-1}    {[} \mathcal{L}_{m} ]_{g^{\mu \nu}} \underset{Eg}{\doteq} - \frac{1}{2}\, T^{(bar)}_{\mu\nu}
 \]
and introduce a  place holder $[ T^{(bar)}]$ in other gauges by formal rescaling (weight $-2$)  the Einstein gauged energy tensor. 
 
Because of
\[
tr\,  {[} \mathcal{L}_{_2\phi} {]}_{g^{\mu\nu}}=0
\]
  and  
\beq
 tr\, {[}\mathcal{L}_H]_{g} = - \mathcal{L}_H + \frac{3\,(\hbar c)^{-1}}{2}(D_{\lambda}D^{\lambda}(\xi \phi)^2)\,\sqrt{|g|}\,    \label{eq trace L-H-g}
\eeq
twice the trace of the Einstein equation (including the  place holder for the baryonic energy tensor) is
\beq
 2\, tr\,[{L}]_g   =  - 2{L}_H + 3\,(\hbar c)^{-1}\xi^2\,D_{\lambda} D^{\lambda}\phi^2  \   - 2{L}_{\phi_2}- {L}_{\phi_3} - 4 {L}_V \; -  tr\, [T^{(bar)}]\,  .   \label{eq trace L-g}
\eeq

\noindent
\textbf{Variation  $\delta \phi$}\\[0.5em]
For the variation of the scalar field  the scale covariance of the expressions in $\phi$ (with values in the respective gauge bundles of the scale group) allows to work with  the Euler expressions in  bundle values, thus avoiding lengthy calculations in derivatives of $\sqrt{|g|}$ \citep[524ff.]{Frankel:Geometry}: 
\beq
 [{L}]_{\phi} = \frac{\delta{L}}{\delta \phi} = \frac{\partial{L}}{\partial \phi} - D_{\mu}\frac{\partial{L}}{\partial(D_{\mu}\phi)} + D_{\mu}D_{\nu}\frac{\partial {L}}{\partial(D_{\mu}D_{\nu}\phi)} \pm \ldots
\eeq
with the Euler equation $[{L}]_{\phi} =0$. Similarly the Euler expression of a vector or tensor field  $Y$ without scaling symmetry (like the scale connection $\varphi$)  can be written as:
\[
 [{L}]_{Y} = \frac{\delta{L}}{\delta Y} = \frac{\partial{L}}{\partial Y} - \nabla(g)_{\mu}\frac{\partial{L}}{\partial(\nabla(g)_{\mu}Y)} \pm  \ldots
\]

In fact, the Weyl geometric compatibility relation  \eqref{eq Dg=0}  implies $D_{\lambda}\sqrt{|g|}=0$, just like the Riemannian compatibility implies $\nabla(g)_{\lambda}\sqrt{|g|}=0$.   Because of this (and of course  $\frac{\delta \sqrt{|g|}}{\delta \phi}=0$)  the Euler expressions of $\delta \phi$ satisfy
$
[\mathcal{L}_X]_{\phi} = [L_X]_{\phi}\sqrt{|g|} \, 
$ 
even for the terms containing derivatives , i.e., $X\in \{\phi_2, \,\phi_3,\,  _2\phi \}$. Therefore 
\[
[\mathcal{L}_X]_{\phi} = 0 \qquad \longleftrightarrow  \qquad [L_X]_{\phi} = 0 \, 
\]

\noindent
Leaving aside coefficients, i.e., using  the  simplified expression
\[
{[}{\tilde{L}}_{\phi_2}] = - D_{\lambda}\phi  D^{\lambda}\phi\, = -  (\partial_{\lambda} - \varphi_{\lambda})\phi\, (\partial^{\lambda} - \varphi^{\lambda})\phi \,  
\]
the  scale covariant Euler expression  is 
 \beqarr  {[}{\tilde{L}}_{\phi_2}]_{\phi} &=& = -D_{\lambda} \frac{\partial {\tilde{L}} }{\partial(D_{\hspace{-0.1em}\lambda} \phi) } =  2\, D_{\lambda} D^{\lambda}\phi \, \\
 &\underset{Eg}{\doteq}& -2 \, (\nabla(g,\varphi)_{\lambda}- 3 \varphi_{\lambda}) (\varphi^{\lambda} \phi_0 )  \nonumber\\
  &\underset{Eg}{\doteq}& -2 \, \big(  \nabla(g)_{\lambda} +\Gamma(\varphi)^{\alpha}_{\alpha \lambda} - 3 \varphi_{\lambda} \big)\varphi^{\lambda} \phi_0 \nonumber \\
    &\underset{Eg}{\doteq}&   -2 \, \big(  \nabla(g)_{\lambda} + \varphi_{\lambda} \big)\varphi^{\lambda} \phi_0\nonumber 
 \eeqarr 
Thus
 \beqarr 
   {[}{L}_{\phi_2}]_{\phi}  &=& \alpha \xi^2\, D_{\lambda}D^{\lambda}\phi\nonumber \\
    &\underset{Eg}{\doteq}&  -\alpha \xi^2  \phi_0\,\big(  \nabla(g)_{\lambda} + \partial_{\lambda}\sigma \big)\partial^{\lambda} \sigma) \, . \label{Euler Lphi2 Eg}
 \eeqarr 
 
 \noindent
For the simplified Lagrangian (leaving out constant factors of  ${[}{L}_{\phi_3}]$) 
\[ {\tilde{L}}_{\phi_3} =    -  \frac{1}{3} \phi^{-2}|D\phi |^3\, =  -\frac{1}{3} \phi^{-2} (s_{\phi}\, D_{\lambda}(\phi)D^{\lambda}(\phi) )^{\frac{3}{2}}\,
\]
the  Euler expression 
is  similarly:
\beqa
\frac{\partial{\tilde{L}}_{\phi_3}}{\partial \phi} &=&  \frac{2}{3}\phi^{-3}\, |D\phi|
^3\,\sqrt{|g|}   = - 2 \phi^{-1} {\tilde{L}}_{\phi_3} \sqrt{|g|} \\
  D_{\lambda} \frac{\partial {\tilde{L}}_{\phi_3} }{\partial(D_{\hspace{-0.1em}\lambda} \phi) } &=&   D_{\lambda}  \big(-\frac{2}{2} \phi^{-2} |D\phi|s_{\phi}\, D^{\lambda}\phi) \big)  \\
   &=&     \big(2 \phi^{-3}D_{\lambda} \phi\, |D\phi|s_{\phi}\, D^{\lambda}\phi) - s_{\phi}\, \phi^{-2} D_{\lambda}(|D\phi|\, D^{\lambda}\phi) \big)  \\
    &=&  -6 \phi^{-1} \mathcal{\tilde{L}}_{\phi_3}  - s_{\phi}\, \phi^{-2} D_{\lambda}(|D\phi|\, D^{\lambda}\phi)   \\
     {[}{\tilde{L}}_{\phi_3}]_{\phi} &=& 4  \phi^{-1} {\tilde{L}}_{\phi_3}  +  s_{\phi}\, \phi^{-2} D_{\lambda}(|D\phi|\, D^{\lambda}\phi)      \\
     &  \underset{Eg}{\doteq} & 4  \phi_0^{-1} {\tilde{L}}_{\phi_3}  - s_{\phi}\, (\nabla(g)_{\lambda} +4 \partial_{\lambda}\sigma - 4 \partial_{\lambda}\sigma)(|\nabla \sigma| \partial^{\lambda}\sigma)  \quad \mbox{(``$-$'' sic; cf. \eqref{eq Gamma "contracted"})}\\
     &  \underset{Eg}{\doteq} &  4 \phi_0^{-1} {\tilde{L}}_{\phi_3}   - s_{\phi}\, (\nabla(g)_{\lambda} (|\nabla \sigma| \partial^{\lambda}\sigma) 
\eeqa
    Thus
\beqarr 
   {[}{L}_{\phi_3}]_{\phi} &=& 4\, \phi^{-1}{L}_{\phi_3}   +  s_{\phi}\,\, \beta \xi^3\, \phi^{-2} D_{\lambda}(|D\phi|\, D^{\lambda}\phi)   \nonumber \\
  &\underset{Eg}{\doteq}&  4 \phi_0^{-1} {L}_{\phi_3}   -  s_{\phi}\,  \beta\,  \xi^3 \, \nabla(g)_{\lambda} (|\nabla \sigma| \partial^{\lambda}\sigma) \, .\label{eq Euler Bek}
\eeqarr 
  For   ${L}_{_2\phi} = (\xi \phi)\,D_{\lambda}D^{\lambda}(\xi \phi)\,A_{\lambda}A^{\lambda} $, up to a constant, the scale covariant variation,
  \[  {[}{{L}}]_{\phi} =  \frac{\partial {{L}} }{\partial \phi} -D_{\lambda} \frac{\partial {{L}} }{\partial(D_{\hspace{-0.1em}\lambda} \phi) } 
 + D_{\lambda}D^{\lambda} \frac{\partial{{L}} }{\partial(D_{\hspace{-0.1em}\lambda}D^{\lambda} \phi)} \,  \] 
is
    \beq  {[}{{L}}_{_2\phi}]_{\phi} =  \, \xi^2 \big(D_{\hspace{-0.1em}\lambda}D^{\lambda} \phi + D_{\hspace{-0.1em}\lambda}D^{\lambda} \phi \big)A_{\lambda}A^{\lambda}\sqrt{|g|} = 2 \, \xi^2 D_{\hspace{-0.1em}\lambda}D^{\lambda} \phi\,A_{\lambda}A^{\lambda} \sqrt{|g|} \, . \label{eq Euler L-X phi}
    \eeq
    
  \noindent
All in all:  
    \beqarr  [{L}_H]_{\phi} &=& \xi (\xi \phi) R(g,\varphi)= 2 \phi^{-1} {L}_H \nonumber \\ 
 {[} {L}_{\phi_2} ]_{\phi} &=&  \alpha \xi^2 D_{\lambda} D^{\lambda}\phi \,  \nonumber  \\
  {[} {L}_{\phi_3} ]_{\phi} &=& 4 \phi^{-1}  {L}_{\phi_3} +   \,s_{\phi}\,\, \beta\, \xi^3  \phi^{-2} D_{\lambda}\big(|D \phi|D^{\lambda}\phi \big) \label{eq Euler phi scale covariant}
  \\
   {[}{{L}}_{_2\phi}]_{\phi}  &=& 2 \, \xi^2 D_{\hspace{-0.1em}\lambda}D^{\lambda} \phi\,A_{\lambda}A^{\lambda}   \nonumber\\
   {[} {L}_{V} ]_{\phi} &=&  4 \phi^{-1} {L}_{V}  \nonumber\\
    {[} {L}_{m} ]_{\phi} &=&  0  \qquad  \big(\mbox{\tiny no coupling of the scalar field to classical  matter} \big) \, , \nonumber 
       \eeqarr 
In Einstein gauge $(g\underset{Eg}{\doteq} g_E$ and $\varphi\underset{Eg}{\doteq} d\sigma$ )   this is:
\beqarr
 {[} {L}_H {]}_{\phi} &\underset{Eg}{\doteq}& (\hbar c)^{-1}\,\xi (\xi \phi_0) R(g,d\sigma)\underset{Eg}{\doteq} 2 \phi_0^{-1} {L}_H(g,d\sigma) \nonumber \\ 
 {[} {L}_{\phi_2} {]}_{\phi} &\underset{Eg}{\doteq}&   -(\hbar c)^{-1}
 \alpha \xi^2  \phi_0\, \Box(g)\sigma  +  2 \phi^{-1}\,{L}_{\phi_2} \hspace{21mm} \mbox{(see \eqref{Euler Lphi2 Eg})}
 \label{eq Euler phi} \\ 
    {[}{L}_{\phi_3}]_{\phi}  &\underset{Eg}{\doteq}&    -  s_{\phi} \beta \, \xi^3\, \nabla(g)_{\lambda} \big(|\nabla \sigma| \partial^{\lambda}\sigma \big) + 4\phi_0^{-1} {L}_{\phi_3} \; \qquad  \qquad  \mbox{(see \eqref{eq Euler Bek})}  \nonumber \\
      {[}{L}_{_2\phi}]_{\phi}  &\underset{Eg}{\doteq}&   - (\hbar c)^{-1} 2 
      \,  \xi^2\phi_0\, \big(\Box(g)\sigma + \partial_{\lambda}\sigma\partial^{\lambda}
      \sigma  \big)    \nonumber \\
   {[} {L}_{V} ]_{\phi} &\underset{Eg}{\doteq}&  4 \phi^{-1} {L}_{V} \,  \nonumber
   \eeqarr     

\subsection{\small Diverse calculations \label{appendix calculations}}

\subsubsection{\small Useful formulas \label{appendix useful formulas}}
\noindent
{\em Selected scale covariant derivatives}:
\begin{subequations}
\beqarr
w(|D\phi|)&=& w(\sqrt{|D_{\nu}\phi D^{\nu}\phi|})=-2  \\
 D_{\lambda}\phi &\underset{Eg}{\doteq}& - \phi_0\, \varphi_{\lambda} \nonumber \\
 D_{\mu}D_{\nu}\phi &=& D_{\mu}\big( (\partial_{\nu} -\varphi_{\nu})\phi\big)= (\nabla_{\mu}- \varphi_{\mu})\big( (\partial_{\nu} -\varphi_{\nu})\phi\big) - \Gamma(\varphi)_{\mu\nu}^{\lambda}(\partial_{\lambda} -\varphi_{\lambda})\phi \nonumber \\
 &\underset{Eg}{\doteq}& \phi_0 \Big((\nabla_{\mu}- \varphi_{\mu})(-\varphi_{\nu}) -\big(\delta^{\lambda}_{\mu}\varphi_{\nu} +\delta^{\lambda}_{\nu}\varphi_{\mu} - g_{\mu\nu}\varphi^{\lambda}  \big)(\partial_{\lambda} -\varphi_{\lambda})\phi \Big) \nonumber \\
 &\underset{Eg}{\doteq}& \phi_0 \Big(-\nabla(g)_{\mu}\varphi_{\nu} + 3 \varphi_{\mu}\varphi_{\nu}- \varphi_{\lambda}\varphi^{\lambda} g_{\mu\nu}  \Big) \label{eq DD phi} \\
 D_{\lambda}D^{\lambda}\phi  &\underset{Eg}{\doteq}&  - \phi_0
\nonumber \big(\nabla(g)_{\lambda}\varphi^{\lambda}+ \varphi_{\lambda}\varphi^{\lambda}   \big)\\
D_{\mu}D_{\nu}\phi^2 &=& 2\, (D_{\mu}\phi D_{\nu}\phi  + \phi\, D_{(\mu}D_{\nu)}\phi) \\
 &\underset{Eg}{\doteq}& 2\phi_0^2\, \big(\varphi_{\mu}\varphi_{\nu}  -\nabla(g)_{\mu}\varphi_{\nu} + 3 \varphi_{\mu}\varphi_{\nu}- \varphi_{\lambda}\varphi^{\lambda} g_{\mu\nu}   \big) \nonumber  \\
  &\underset{Eg}{\doteq}& 2\phi_0^2\, \big(4 \varphi_{\mu}\varphi_{\nu}  -\nabla(g)_{(\mu}\varphi_{\nu)} - \varphi_{\lambda}\varphi^{\lambda} g_{\mu\nu}   \big) \\
  D_{\lambda}D^{\lambda}\phi^2 &=& 2\,(D_{\lambda}\phi D^{\lambda}\phi  + \phi\, D_{\lambda}D^{\lambda}\phi) \label{eq D-lambda-square phi-square} \\
    &\underset{Eg}{\doteq}& 2\phi_0^2\, \big(4 \varphi_{\lambda}\varphi^{\lambda}  -\nabla(g)_{\lambda}\varphi^{\lambda} -4  \varphi_{\lambda}\varphi^{\lambda}   \big) = - 2\phi_0^2\,\,\nabla(g)_{\lambda}\varphi^{\lambda} \nonumber \\
    D_{\lambda}D^{\lambda}\phi^2\, g_{\mu\nu}  - D_{\mu}D_{\nu}\phi^2  &\underset{Eg}{\doteq}&  2\phi_0^2\Big(\big( \partial_{\lambda}\sigma\partial^{\lambda}\sigma -\nabla(g)_{\lambda}\partial^{\lambda}\sigma \big) g_{\mu\nu} - 4\, \partial_{\mu}\sigma  \partial_{\nu}\sigma + \nabla(g)_{(\mu}\partial_{\nu)} \sigma  \Big) \label{eq Theta-phi-H in Einstein gauge}
\eeqarr
\end{subequations}

\noindent
{\em Diverse derivative expressions}
\beqarr
\partial_{\lambda} \sqrt{|g|} &=& \frac{1}{2}\frac{\partial_{\lambda}|g|}{\sqrt{|g|}} =  \frac{1}{2}\frac{\partial_\lambda |g|}{|g|}  \sqrt{|g|} =  \frac{1}{2} \partial_{\lambda} \ln |g| \sqrt{|g|}= g^{\mu \nu} \partial_{\lambda}g_{\mu \nu} \sqrt{|g|} \nonumber  \\
\partial_{\lambda} \ln |g| &=& g^{\mu \nu} \partial_{\lambda}g_{\mu \nu} \quad \qquad \mbox{\citep[p. 124, eq. (134)]{Tann:Diss}}  \nonumber \\
\Gamma(g)^{\lambda}_{\lambda \alpha} &=& \frac{\partial_{\alpha}\sqrt{|g|} }{\sqrt{|g|}} \qquad \qquad \mbox{\citep[p. 101, eq (3.33)]{Carroll:Spacetime}} \nonumber \\
 \Gamma(\varphi)_{\lambda \nu}^{\lambda} &=& 4 \varphi_{\nu} \label{eq Gamma "contracted"} \\
 \partial_{\lambda} \big(X^{\lambda} \sqrt{|g|}  \big)  &=& \nabla(g)_{\lambda}X^{\lambda}\,\sqrt{|g|}  \nonumber  
\eeqarr

\subsubsection{\small Centrally symmetric metric \label{subsection appendix central symmetric}}

\subsubsection*{\small Milgrom  equation}
For the calculation of the Milgrom operator 
$
 \nabla(g)_{\lambda}(|\nabla \sigma |\partial^{\lambda}\sigma \big) 
$,  the l.h.s. of  \eqref{eq core Milgrom equation} 
 in the case of a
centrally symmetric metric  with area radius  \eqref{eq central symmetric metric},   one needs
(in the following $\Gamma$ stands for  $\Gamma(g)$)
\[ \Gamma^1_{00} =  \frac{a'}{2b}, \quad  \Gamma^1_{11}= \frac{b'}{2b}, \quad  g^{jj}\Gamma^1_{jj}= - \frac{1}{rb} \;\; \mbox{(without  summation) for}\; j= 2, 3 \;    \mbox{}
\]
and for  the d'Alembert operator 
additionally
\[ \Gamma^0_{01}=\frac{a'}{2a}\, \quad \Gamma^2_{21}=\Gamma^3_{31}= \frac{1}{r} \, .
\]

\noindent
The d'Alembert operator is here (abbreviating  $\underset{Eg}{\doteq}
$  by $=$) 
\beq \Box(g) \sigma = \frac{\sigma''}{b} + \sigma' \big(\, \frac{1}{2b}(\frac{a'}{a} + \frac{4}{r}) - \frac{b'}{2b^2}   \big) \, 
\eeq 
and the Milgrom operator
\beq
  b^{-\frac{1}{2}} |\sigma'|\Big(\frac{2 \sigma ''}{b} + \sigma' \big(\frac{a'}{2ab} - \frac{b'}{b^2} + \frac{2}{r b} \big)  \Big) \, .  \label{eq central symmetric Milgrom operator}
\eeq
The scalar field equation in baryonic vacuum is  satisfied for
\[ \sigma'' = \sigma' \big(\frac{b'}{2b} - \frac{a'}{4a} - \frac{1}{r} \big) \qquad \mbox{or} \quad \sigma' = 0 \, .
\]
Integration of the  first condition results in 
\beq \sigma' = b^{\frac{1}{2}} a^{-\frac{1}{4}}\, \frac{c_1}{r} \, , \qquad 
\sigma''= -  b^{\frac{1}{2}} a^{-\frac{1}{4}}\, \frac{c_1}{r^2} + \partial_r  \big(b^{\frac{1}{2}} a^{-\frac{1}{4}}\big)\, \frac{c_1}{r}  \; , \label{eq sigma on shell}
\eeq
which looks  similar to  the flat case of MOND. One should, however, not take the same   constant $c_1= \sqrt{a_1 M}$ as there. From \eqref{eq Poisson equation Newton approximation} and \eqref{eq Newtonian mass equivalent phi 2} we can conclude that in the relativistic approach the constant has  to be set 
\[
 c_1= \frac{1}{2}\sqrt{a_1 M}\, ,
\]
in order to get a comparable model dynamic.

The energy component of the r.h.s of the Einstein equation without baryonic/classical matter is given by \eqref{eq Theta-phi-bar}. 
For metrics not too far away from Schwarzschild  $\Theta^{(\sigma)}$ is strongly dominated by the second order terms.  We  therefore replace  \eqref{eq Theta-phi-bar} on the r.h.s. of the Einstein equation with its   abbreviated form 
\eqref{eq Theta-bar}
\[ \hat{\Theta}^{(\sigma)}_{\mu\nu} =  \Box(g)\sigma\, (g_{\mu\nu} + 2 a_{\mu}a{_\nu}) - 2\, \nabla(g)_{(\mu}\sigma_{\nu)}  \, .
\]
 where $A=(1,0,0,0)$.

\subsection{\small Dimensional conventions \label{appendix dimensional considerations}}

Denote the physical dimension of  a quantity $X$ by square brackets $[X ]$, and  the Weyl weight by double square brackets $[[X]]=w(X)$.  In this paper the  following conventions are used:
\\
\noindent
Elementary dimensions (like in the SI conventions) 
 {\em length} $  \mathtt{L}$, {\em time} $\mathtt{T}$, {\em energy} $\mathtt{E}$ 
 have the Weyl weights $[[L]]= [[T]]=1, \; \; [[E]]=-1$. 
Universal constants 
 $c$ {\em speed of light}, $\hbar$ {\em Planck constant} with    $[c]=\mathtt{ L T^{-1}}$,  $[\hbar]= \mathtt{E\, T}$ 
 have the  Weyl weights $[[c]]=0$,  $[[\hbar]]= 0$ (Weyl invariant).

 Derived dimensions are built from them. 
Basic ones are 
{\em mass} $ [m] =\mathtt{M} = \mathtt{E\, L^{-2} T^{2}}$,
 {\em velocity} $[v]= \mathtt{LT^{-1}}$, {\em acceleration} $[a]= \mathtt{LT^{-2}}$ etc. 
($[[m]]=-1, \; [[v]]=0,\; [[a]]=-1$). \\
Dimensional constants: \\
{\em Newton constant} $[G]= \mathtt{L^3 \ T^{-2}\,M^{-1}}$, $[[G]]=2$, \\
{\em Einstein gravitational  constant} $\varkappa = G\, c^{-4}$, $[\varkappa]=\mathtt{L^{-1}\, T^2\, M^{-1}} = \mathtt{L\,E^{-1} } $, $[[\varkappa]]=2$,\\
 {\em Schwarzschild mass factor} $M= m\, G\, c^{-2}$, $[M]= \mathtt{L}$, $[[M]]=1$.\\[-0.8em]

\noindent
\textbf{Dimensional conventions in classical physics and SR}

\noindent
In {\em Euclidean-Galilean or special relativistic space-time} 
 coordinates $x_i, t$ are often treated as dimensional quantities (because determined by measurements):
\[ [x_i] = L \, , \qquad [t] = T \, , \qquad \quad  ([[x_i]]= [[t]]=1)
\]
Partial derivatives then change  dimensions (and Weyl weights):\\
$[\partial_i] = \mathtt{L^{-1}}, \; [\partial_t] = \mathtt{T^{-1}}$,
$[\Delta]= \mathtt{L^{-2}}$ etc. \quad  ($[[\partial_i]] = [[\partial_t]] = -1, [[\Delta]]=-2$).\\[-0.8em]

The Newton potential $\Phi_N$  is of dimension $[\Phi_N]= \mathtt{L^2T^{-2}}$. For a mass density with $[\rho_m] = \mathtt{M\, L^{-3}}$ and $[\Delta]=L^{-2}$ the Poisson equation is dimensionally consistent with $[\Delta\, \Phi_N]=[ G\, \rho_m]= \mathtt{T^{-2}}$; the Newtonian tidal tensor has dimension $[\tau]= \mathtt{T^{-2}}$ 
etc.\\[-0.8em]

\noindent
\textbf{Dimensional conventions  in GR}\\
In the case of a {\em Lorentzian (or Weyl geometric) spacetime} this is different; coordinates are non-metrical and should thus be treated as dimension-less functions on the manifold.\footnote{In Weyl's words: ``The coordinates which are arbitrarily projected into the world have to be considered as dimension-less numbers, the determination of  $ds^2$ presupposes an arbitrarily chosen unit of measurement, \ldots'' \citep[p. 232, my transl.]{Weyl:RZM5}, see also the commentary in \citep[p. 629ff.]{Giulini/Scholz:RZM}. 
This has to be kept in mind even in cases where coordinates  {\em mimick  metrical relations} to a certain degree  (e.g. Schwarzschild metric, Robertson Walker metric, etc.). } 
It is the physical fields  on the manifold, which carry dimensions; among them 
 the coefficients of the Riemannian {\em metric coefficients} $ [g_{\mu\nu}]=\mathtt{L^2}, \; [g^{\mu\nu}]=\mathtt{L^{-2}}$ and  $[[g_{\mu\nu}]]=2,\, [[g^{\mu\nu}]]=-2$.  
 In the present paper the {\em scalar field} is  endowed with the dimension $[\phi]=\mathtt{E}$ (alternative conventions might be $\mathtt{L^{-1}}$, respectively $\mathtt{T^{-1}}$;  all of the same Weyl weight $-1$).\\  
Dimensionless quantities are: the coefficients of the 
 {\em affine connection}, in abbreviated notation $[\Gamma]=1$,\footnote{The dimensional factors $\mathtt{L^2}, \, \mathtt{L^{-2}}$ in  terms of the form $g^{\mu \alpha} \partial_{\alpha}g_{\nu \lambda} $ cancel.}
the Riemann and Ricci curvatures $[Riem]=1, \, [Ric]=1$ ($[[Riem]]=[[Ric]]=0$) and the {\em Einstein tensor} $[Ric- \frac{1}{2}R\, g]=1$ (in the Weyl geometric case $[[Ric- \frac{1}{2}R\, g]]=0$). Derivatives (partial, covariant, scale covariant) do not  change dimensions. \\
Curvature quantities which carry dimensions are the  scalar curvature $[R]=L^{-2}$ ($[[R]]=-2$) and the sectional curvatures $\kappa(u,v)$ with regard to tangent directions $u, v$,  $[\kappa(u,v)]=L^{-2}$.\\
[-0.8em]
 
 The energy momentum tensor $T^{\mu}_{\; \nu}$ has the dimension of a spatial energy density, $[T^{\mu}_{\;\; \nu}]= \mathtt{E\, L^{-3}}$; thus 
 $[T_{\mu \nu}]= \mathtt{ E\, L^{-1}}$; in particular the energy density $\rho_e = T^0_{\;\;0}$  with $[\rho_e]= \mathtt{EL^{-3}}$ implies that  $T_{00}= \rho = g_{00}\,\rho_e$ has dimension $[\rho]=\mathtt{EL^{-1}}$. This makes  both sides of the  Einstein equation dimensionless -- but the transition to the Newton approximation is  dimensionally  tricky.\\[-0.5em]
 
 \noindent
 \textbf{Dimensions in the Newton approximation}\\
  A  fluid with relativistically negligible pressure (dust) with flow covector field $u=(u_{\mu})$ of  weight $1$ and dimension $\mathtt{L}$ has energy momentum $T_{\mu\nu} =  \rho_m \, u_{\mu}u_{\nu}$, where  in the rest system $(u_{\mu})= (1,0,0,0)$. Then $[T_{\mu \nu}]= \mathtt{E L^{-3}L^2}= \mathtt{EL^{-1}}$ as it should be. 
  The energy component of  the Einstein equation is 
  \beq R_{00} \approx  8 \pi \varkappa \,(T_{00}-\frac{1}{2}\, tr\, T) = 4 \pi \varkappa\, \rho \, . \label{eq energy component of weak field approximation}
  \eeq
  Both sides are dimensionless (in our convention). 
The first order approximation of $R_{00}$ for the weak field \eqref{eq weak field metric} is like in  \eqref{eq Ri0j0 weak field} 
\[
R_{00} \approx - \frac{1}{2} \delta^{ij}\, \partial_i\partial_j h_{00}
\]
with dimensions $[h_{00}]=1$, $[\partial_i]= 1$  
(they are dimension-less).  This and \eqref{eq energy component of weak field approximation}  multiplied with $-2 c^2$ gives
\[
c^2\, \nabla^2 h_{00} \approx - (8\pi\, G)\, c^{-2}\, \rho
\]
(both sides of dimension $\mathtt{L^2T^{-2}}$ in GR conventions). With  the relation between $h_{00}$ and the Newton potential  $c^2h_{00}=-2\Phi$ \eqref{eq Phi -- h00}  {\em and  
a reinterpretation of the r.h.s. factor} $c^{-2}\rho$ {\em  as a 
 mass density proper},  $\rho_m= c^{-2}\, \rho_e$ with $[\rho_m]= \mathtt{ML^{-3}}$,
 the bridge between the energy component of the Einstein equation \eqref{eq energy component of weak field approximation} and the Newtonian Poisson equation is established,
\[
\nabla^2 \Phi \approx 4\pi G\, \rho_m \, . 
\]
and is dimensionally consistent in the flat space conventions, $[\nabla^2 \Phi]= \mathtt{L^{-2} L^2 T^{-2}}= \mathtt{T^{-2}}$. \\[-0.8em]

\noindent
\textbf{Dimensions of Lagrangians and of dynamical equations }\\
Dimensions of Lagrange densities \eqref{eq Lagrangian phi-2} etc.   $[\mathcal{L}_X]= E L$.\\
Both sides of the Einstein equation are dimension-less, with  $[T_{\mu\nu}]=E L^{-1}, \, [\varkappa] = L E^{-1}$ as above,  $[\Theta^{(X)}_{\mu\nu}] =1$.\\
The dimension of the full Milgrom equation \eqref{eq full Milgrom equation} is $L^{-3}$,  
 while for the deep MOND equation \eqref{eq flat space Milgrom equation} the dimension  in flat space conventions  is $L T^{-4}$  etc.\\[-0.3em]

\vspace{3em}

\addcontentsline{toc}{section}{Bibliographie}
\scriptsize


\end{document}